\documentclass[12pt]{article}
\usepackage{epsfig}
\input epsf.sty
\newcommand{\la}[1]{\label{#1}}
 
\newcommand{\be}{\begin{equation}}
\newcommand{\ee}{\end{equation}}
\newcommand{\ba}{\begin{eqnarray}}
\newcommand{\ea}{\end{eqnarray}}
\newcommand{\bi}{\begin{itemize}}
\newcommand{\ei}{\end{itemize}}
\newcommand{\rmi}[1]{{\mbox{\scriptsize #1}}}
\newcommand{\nr}[1]{(\ref{#1})}
\newcommand{\tr}{{\rm Tr\,}}
\newcommand{\re}{\mathop{\rm Re}}

\newcommand{\nn}{\nonumber \\}
\newcommand{\fr}[2]{{\frac{#1}{#2}}}

\renewcommand{\vec}[1]{{\bf #1}}

\newcommand{\RR}{{\rm I\kern -.2em  R}} 
\newcommand{\eq}{Eq.~}
\newcommand{\eqs}{Eqs.~}
\newcommand{\fig}{Fig.~}

\newcommand{\se}{Sec.~}

\def\lsi{\raise0.3ex\hbox{$<$\kern-0.75em\raise-1.1ex\hbox{$\sim$}}}
\def\gsi{\raise0.3ex\hbox{$>$\kern-0.75em\raise-1.1ex\hbox{$\sim$}}}

\newcommand{\gsim}{\mathop{\gsi}}

\makeatletter \@addtoreset{equation}{section} \makeatother
\renewcommand{\theequation}{\arabic{section}.\arabic{equation}}
\makeatletter
\renewcommand\section{\@startsection {section}{1}{\z@}%
                                   {-5.5ex \@plus -1ex \@minus -.2ex}
                                   {2.3ex \@plus.2ex}%
                                   {\normalfont\large\bfseries}}
\renewcommand\subsection{\@startsection{subsection}{2}{\z@}%
                                     {-3.25ex\@plus -1ex \@minus -.2ex}%
                                     {1.5ex \@plus .2ex}%
                                     {\normalfont\normalsize\bfseries}}
\renewcommand\thesection {\@arabic\c@section}
\renewcommand\thesubsection   {\thesection.\@arabic\c@subsection}
\renewcommand{\@seccntformat}[1]{%
\csname the#1\endcsname.\hspace{1.0em}}
\makeatother

\begin{document}
 
\begin{titlepage}
\begin{flushright}
CERN-TH/2002-319\\
HIP-2002-55/TH\\
NORDITA-2002-69 HE\\
UNIL-IPT-02-11\\
hep-ph/0211149\\
\end{flushright}
\begin{centering}
\vfill
 
{\bf LOCALISATION AND MASS GENERATION\\ FOR NON-ABELIAN GAUGE FIELDS}

\vspace{0.8cm}
 
M. Laine$^{\rm a}$, 
H.B. Meyer$^{\rm b}$, 
K. Rummukainen$^{\rm c,d}$, 
M. Shaposhnikov$^{\rm e}$ 

\vspace{0.3cm}

{\em $^{\rm a}$%
Theory Division, CERN, CH-1211 Geneva 23,
Switzerland\\}

\vspace{0.3cm}

{\em $^{\rm b}$%
Theoretical Physics, University of Oxford, 
1 Keble Road, Oxford, OX1 3NP, UK\\} 

\vspace{0.3cm}

{\em $^{\rm c}$%
Department of Physics,
P.O.Box 64, FIN-00014 University of Helsinki, Finland\\}

\vspace{0.3cm}

{\em $^{\rm d}$%
NORDITA, Blegdamsvej 17,
DK-2100 Copenhagen \O, Denmark\\}

\vspace{0.3cm}

{\em $^{\rm e}$%
Institute of Theoretical Physics, 
University of Lausanne,\\ 
BSP-Dorigny, CH-1015 Lausanne, Switzerland}

\vspace*{0.8cm}
 
\end{centering}
 
\noindent 
It has been suggested recently that in the presence of suitably ``warped" 
extra dimensions,  the low-energy limit of pure gauge field theory may 
contain massive elementary vector bosons localised on a ``brane'', but no 
elementary Higgs scalars. We provide non-perturbative evidence 
in favour of this conjecture through numerical lattice measurements
of the static quark--antiquark force of pure SU(2) gauge theory 
in three dimensions, of which one is warped. 
We consider also warpings leading to massless localised vector bosons, 
and again find evidence supporting the perturbative prediction, even 
though the gauge coupling diverges far from the brane in this case. 
\vfill
\noindent
 

\vspace*{1cm}
 
\noindent
CERN-TH/2002-319\\
January 2003 

\vfill

\end{titlepage}

\setcounter{footnote}{0}

\section{Introduction}
\la{se:intro}

In standard Kaluza--Klein dimensional reduction
of  pure gauge theory, the original (say,
five-dimensional) theory has an effective description in
terms of a four-dimensional theory, whose lightest degrees of freedom
are in the Coulomb or confinement phase (depending on the group), and
have  a wave function spread out evenly in the fifth dimension. It
has recently been demonstrated~\cite{st}  that if the fifth dimension
is suitably ``warped'', this pattern could change qualitatively, at
least in the Abelian case: the low-energy dynamics can still be 
four-dimensional, but now with {\em massive} elementary vector bosons, 
and with a {\em localised}  wave function along the extra
dimension.  Thus, extra dimensions could potentially provide an
alternative for the Higgs mechanism.

While there is no doubt about the viability of this mechanism in the
Abelian case, where all computations can be carried out
analytically,  things are more complicated in a non-Abelian theory.
For a specific choice of the warp factor the low-energy
effective action looks much like a four-dimensional gauge theory, but with a
gauge non-invariant mass term. For an Abelian case, this is still a
renormalisable theory, whereas
for non-Abelian groups it is in general 
not (see, e.g., ref.~\cite{Bardeen:1978cz}). This
means that the heavier modes cannot decouple from the low-energy
dynamics. The hope is that they might nevertheless only introduce
small contributions, like higher order operators do in chiral
perturbation theory, but this has  so far not been demonstrated
explicitly. 

Another way to express the problem is that 
a gauge theory with a mass term ``introduced by
hand" may be considered the infinite Higgs-mass limit of
a gauge-Higgs theory with spontaneous symmetry breaking
\cite{Bardeen:1978cz}--\cite{Appelquist:1980vg}, 
and is therefore strongly coupled, at
energies of the order of the vector boson mass. Therefore
the viability of perturbation theory must again be
checked by non-perturbative means. 

There are also other types of warp factors, 
discussed in connection 
with the localisation of
gravity~\cite{rs} and gauge fields on a brane, which lead 
again to a lower dimensional effective theory, but this time with massless 
vector bosons (see, e.g., \cite{Dvali:1997xe}--\cite{Randjbar-Daemi:2002pq}).
This requires asymptotically small warp factors (or, in other terms, 
large gauge couplings) far from the brane~\cite{st}.
As in the previous case, the validity of perturbation theory 
is then in question. Some aspects related
to this mechanism were already studied with numerical methods in~\cite{dfkk}.

The purpose of the present paper is to study the issue of strong
coupling with lattice  Monte Carlo simulations. To simplify the
analysis, we would like to separate the problem of
non-renormalisability of the higher dimensional original
gauge theory from the problem of a large coupling constant far from a
brane. To this end one can study a compactification from four
dimensions (4d) to three  dimensions (3d), or even three dimensions to two
dimensions (2d). For the practical reasons that less computer time is
required, and some exact results are available in 2d physics,  we
choose here the latter case. Nevertheless, we should expect the main
features to carry over to higher dimensions, as well.  

The outline of the paper is the following.  We review some basic
aspects of the mechanism in~\se\ref{se:review}. We introduce our
observables and determine their behaviour in the Abelian case
in~\se\ref{se:abelian}. The lattice formulation is  presented
in~\se\ref{se:nonabelian}, and numerical results for the 
Abelian and non-Abelian cases, in~\se\ref{se:lattice}. 
We conclude in~\se\ref{se:concl}. 
Some technical details are discussed in the Appendices.

\section{The mechanism in review}
\la{se:review}

We start by reviewing the basic properties of the mechanism 
introduced in~\cite{st}, in the Abelian case. 
The Euclidean continuum action is 
\be
 S_E^{(d+1)} = \int\! {\rm d}^d x\, \int\! {\rm d} z\,  \Delta(z) 
 \fr14 F_{\tilde\mu\tilde\nu} F_{\tilde\mu\tilde\nu} \;, 
 \la{action}
\ee
where $F_{\tilde\mu\tilde\nu} = \partial_{\tilde\mu} A_{\tilde\nu}
-   \partial_{\tilde\nu} A_{\tilde\mu}$, and $\tilde \mu =
1,...,d+1$,  $z \equiv x_{d+1}$. An index without a tilde runs as
$\mu = 1,...,d$.  The function $\Delta(z) > 0$ breaks the
($d+1$)-dimensional Lorentz invariance.  We however assume the
special breaking pattern  that terms containing $F_{\mu\nu}$, $F_{\mu
z}$  are multiplied by the same function. We also take  $\Delta(z)$
to be an even function of $z$, and refer sometimes to the 
plane $z=0$ as the ``brane''.

We choose units such that \eq\nr{action} should roughly correspond to
an effective $d$-dimensional  action of the form 
\be
 S_E^{(d+1)} \sim \int\! {\rm d}^d x\, \frac{1}{4 g^2}
 F_{\mu\nu} F_{\mu\nu} + ...\; , 
\ee
where $g$ is the gauge coupling (which of course plays no dynamical
role in the non-interacting Abelian case). Thus, 
\be
 [A_\mu] = \mbox{GeV}, \quad
 [F_{\mu\nu}] = \mbox{GeV}^2, \quad
 [g^2] = \mbox{GeV}^{4-d}, \quad
 [\int_z \Delta(z)] = \Bigl[\frac{1}{g^2}\Bigr]. 
\ee 

To proceed,  we assume that one can make the gauge choice $A_z = 0$, 
without introducing any singularities. 
It should be noted, however, that this may not always
be the case in a strict sense, in a non-Abelian theory. 
If for instance the extent of the
$z$-direction and $\int_z \Delta(z)$ are finite, like at  finite
temperatures,  then $A_z$ behaves effectively  like a dynamical
adjoint-charged scalar field related to the global  symmetries of the
system, which can even get spontaneously 
broken~\cite{gpy,nw,hosotani} (for a recent study in the context
of extra dimensions, see~\cite{kal}). For the purposes of this
Section, though, this possibility can be ignored.  
It should perhaps be stressed that all the observables 
to be introduced later on, as well as the main lattice 
simulations carried out, are explicitly gauge invariant, 
so that our conclusions are by construction based on data
which is independent of the gauge choice. 

We now carry out a mode decomposition of the functional dependence 
of the fields on the $z$-coordinate, 
\be
 A_\mu(x,z) = \sum_n A_\mu^n(x) \psi_n(z).
\ee
Units are chosen such that 
\be
 [\psi_n] = [g] = \mbox{GeV}^{(4-d)/2}.
\ee
The real functions $\psi_n(z)$ are assumed to satisfy the second order
Sturm--Liouville linear differential equation,
\be
 -\frac{1}{\Delta(z)} \Bigl[\Delta(z) \psi'_n(z) \Bigr]' = m_n^2 \psi_n(z). 
 \la{diffeq}
\ee
Here $m_n^2$ are real, because the differential operator is
Hermitean.  They turn out also to be non-negative. We denote 
the mode constant in $z$ (whether normalisable or not in infinite volume)
by the index $n \equiv c$, while general normalisable states with
non-negative masses are labeled by $n\ge 0$, with even (odd) indices
denoting states symmetric (anti-symmetric) in $z\to -z$. 
Explicit solutions  of~\eq\nr{diffeq} for various
$\Delta(z)$ are discussed in~\ref{app:cases}.
Note that if the constant mode is normalisable in infinite volume, 
then the indices $n=0$ and $c$ refer to one and the same mode. 

Together with the normalisation condition
\be
 \int_z \Delta(z) \psi_m(z) \psi_n(z) \equiv \delta_{mn}, 
\ee
\eq\nr{diffeq} guarantees that 
\be
 \int_z \Delta(z) \psi'_m(z) \psi'_n(z) = m_n^2 \delta_{mn}. 
\ee
Note also that the completeness relation can be written as 
\be
\sum_n \psi_n(z) \psi_n(z') = \Delta^{-1}(z) \delta(z-z').
\ee 
The quadratic part of the action then becomes
\be
 S_E^{(d+1)} = 
 \int {\rm d}^dx \sum_{n \ge 0} \Bigl( 
 \fr14 F^n_{\mu\nu} F^n_{\mu\nu} + \fr12 m_n^2 A_\mu^n A_\mu^n
 \Bigr).
 \la{n_action} 
\ee
If the lowest mass is zero or much smaller than the masses of higher 
modes, we have an effective $d$-dimensional field theory at low energies:
it is described by the term with $n=0$ (or $n=c$) in~\eq\nr{n_action}. 

Now, if the extent of the $z$-direction is finite
and $\Delta(z)$ is regular, or if
\be
 \int_{-\infty}^{\infty} \! {\rm d} z \, \Delta(z) <\infty , 
 \la{cond}
\ee
then \eq\nr{diffeq} clearly  has 
a normalisable zero mode solution, with $\psi_0(z)=\psi_c$
constant and $m_0^2 = m_c^2 = 0$. The normalised form of this solution is 
\be
 \psi_c = \frac{1}{\sqrt{\int_z \Delta(z)}}. 
\ee 
Then the low-energy effective theory is simply 
a standard pure gauge theory. The condition
\eq\nr{cond} implies that $\lim_{z\to\infty}\Delta(z)=0$ and,
therefore, that the effective $d+1$ dimensional gauge coupling
$\Delta^{-1}(z)$ is
large far from the brane. In the non-Abelian case
this fact may, in principle, invalidate the
perturbative arguments just presented, and thus provides a
motivation for a lattice study.

If the condition in \eq\nr{cond} is not satisfied, 
then the constant mode $\psi_c$ 
effectively decouples (since $\psi_c\to 0$); $m_0 \neq 0$; and we
have massive vector bosons without any scalar particles. 
Furthermore, a mass
hierarchy $m_0^2 \ll m_1^2$ can be achieved with some choices
of warp factors (see  \ref{app:cases} and ref.~\cite{st}), provided that
$\Delta(0)/\Delta(z_0)\gg 1$, where $z_0$ is a point where $\Delta(z)$
reaches its minimum value. Thus, a large mass ratio again only 
appears if the effective
higher-dimensional gauge coupling $\Delta^{-1}(z)$
is large, but now at a finite
distance $z_0$ from the brane.

Another subtle point with the case $m_0 \neq 0$ is that the low
energy action is seemingly not gauge invariant (see \cite{st} for a
discussion of gauge transformations). In the Abelian case the theory
is nevertheless renormalisable, even if some  interactions were added
(see, e.g.,~\cite{zj}). This is no longer true for non-Abelian
theories, and the question appears whether the higher lying modes
decouple or not.

\section{Static force in the continuum}
\la{se:abelian}

In order to distinguish the two different regimes (with and without
the massless vector mode $\psi_c$) 
we shall employ the standard order parameter
for confinement, the static force between two heavy test charges
in the fundamental  representation. We measure the force
at a fixed $z$; for actual mechanisms for the localisation of scalars
and  fermions in the vicinity of $z=0$ 
see, e.g.,~\cite{Rubakov:2001kp} and references therein.

For now, we shall restrict to $d=2$, the case we have actually 
studied with lattice simulations. We consider a rectangular area
with $(\Delta x_1,\Delta x_2) \equiv (r,t)$,
and define a Wilson loop around the rectangle, 
\ba
 W(r,t;z) & = &  \Bigl\langle 
 \re \tr {\cal P} \exp(i \oint A_\mu(x,z) \, {\rm d} x_\mu) 
 \Bigr\rangle \nn 
 & = & \Bigl\langle 
 \re \tr {\cal P} \exp(i \sum_n \psi_n(z) \oint A^n_\mu(x) \, {\rm d} x_\mu) 
 \Bigr\rangle. \la{defW}
\ea
The static potential can then be obtained as usual, 
\be
 V(r;z) = - \lim_{t \to \infty} \frac{1}{t} \ln W(r,t;z). \la{defV} 
\ee
A lowest order computation gives 
\be
 V(r;z) = - \sum_n \psi_n^2(z) \int \frac{{\rm d} p}{2\pi} 
 \frac{e^{i p r} - 1}{p^2 + m_n^2} = 
 \sum_n  \frac{\psi_n^2(z)}{2 m_n} \Bigl( 1 - e^{-m_n r} \Bigr)\;. 
 \la{lo_pot}
\ee

The static potential, itself, is of course not a physical observable. 
Depending on the spectrum $m_n$, its absolute value can be ultraviolet 
divergent, and in any case sensitive to ultraviolet physics. Therefore
we rather address its derivative, the force $F(r;z)$,
\be 
 F(r;z) \equiv \frac{ \partial V(r;z)}{\partial r} \;.
 \la{force}
\ee
According to~\eq\nr{lo_pot}, 
\be
 F(r;z) = 
 \sum_n \fr12 \psi_n^2(z) e^{-m_n r} \;. 
 \la{Fr}
\ee
We note from~\eq\nr{Fr} that an external source couples 
to the mode $n$ via $g_n^\rmi{ext}\equiv \psi_n(z)$.

The signatures expected from $F(r;z)$ can  thus be summarised
as follows. In the case that the zero mode exists, $m_0 = m_c = 0$, the 
force should approach a constant at large~$r$, 
\be
 F(r;z) \to \fr12 \psi_c^2, 
 \la{linear}
\ee
because massive modes give contributions screened at distances
$r\gsim 1/m_n$. On the other hand, in the case of interest to us 
where $m_0 \neq 0,~m_0 \ll m_1$,
and the zero mode decouples ($\psi_c\to 0$), we expect 
\be
 F(r;z) \approx 
 \fr12 \psi_0^2(z)\, e^{-m_0 r} \;. 
 \la{fundmode}
\ee
It is thus our objective to show that the force does get screened, 
but only on large distances, as determined by $1/m_0$. 

While we focus on the force in this paper, 
a behaviour qualitatively very similar can, particularly 
in the Abelian case, be obtained from various correlators of
local gauge invariant operators. For completeness, 
we discuss one example in~\ref{app:correlations}.

So far we have discussed the potential in the Abelian theory. 
In the non-Abelian case, the self- and cross-interactions between 
modes make obviously a fully analytic computation impossible.
However, if dimensional reduction indeed takes place then,  
as discussed in~\ref{app:tension}, 
the only change in the long-distance force
is a colour factor, the quadratic Casimir of the fundamental 
representation,  $C_A = ({N_c^2 - 1})/({2 N_c})$:
\be
 F(r;z) \to C_A F(r;z)  \;. 
 \la{CA}
\ee
This simple relation, which allows us to directly compare 
the asymptotic non-Abelian force with the Abelian one, 
is obviously specific to 2d physics. 

In the non-Abelian case, it is useful to also  
define couplings characterising the cubic and quartic 
self-interactions of the fundamental mode.
Let us introduce
\begin{equation}
 g_3\equiv \int_z \Delta(z) \psi_0^3(z), 
 \qquad 
 g_4^2 \equiv \int_z \Delta(z) \psi_0^4(z) \;,
\end{equation}
and construct the dimensionless quantities
\be
 \alpha_3\equiv \frac{g_3}{g_0^\rmi{ext}}, 
 \qquad 
 \alpha_4\equiv \frac{g_4^2}{[g_0^\rmi{ext}]^2}.
 \la{al}
\ee
In order for the low-energy effective theory 
to be ``close'' to a gauge theory,  
these numbers had better be close to unity.  In
particular, if the zero mode is normalisable and therefore
$\psi_0=\psi_c$, we have exactly $\alpha_3=\alpha_4=1$. In the
opposite case of $m_0 \neq 0$,
we have $\alpha_3 \neq 1$ and $\alpha_4\neq 1$.
Thus, the breaking  of gauge invariance in the low-energy sector
manifests itself both  through an effective mass term
in~\eq\nr{n_action}, and through
non-universal self-interactions which differ from the coupling of
the modes to external sources.

\section{Static force on the lattice}
\la{se:nonabelian}
\la{se:2d}

As mentioned above, in the non-Abelian case the heavier modes cannot
decouple, because they are needed to guarantee renormalisability. It
is therefore not obvious how well the analytical estimates
presented in~\se\ref{se:abelian} really hold. We
will hence study that system with simple numerical lattice Monte
Carlo simulations. 

In fact, to account properly for finite size and finite lattice
spacing effects, we will carry out small scale simulations
for the Abelian  system, as well. Thus, we can directly compare the
two sets of data, with similar volumes and lattice spacings. This may
be useful because the ``sharp'' and ``smooth'' weight functions to be
introduced contain  a small scale hierarchy, which tends to lead
either to finite size or finite lattice spacing effects in lattice
simulations.  Still, both sets of results turn out  in most cases to
remain close to the analytic continuum estimates.

In the Abelian case, we discretise the action in~\eq\nr{action} by using
the so-called non-compact formulation: 
\be
 S_E^{(d+1)} = \sum_z \beta_G(z) \sum_x \sum_{\tilde\mu < \tilde\nu}
 \fr12\alpha_{\tilde\mu\tilde\nu}^2,  
 \la{ab_latt_action}
\ee
where 
$\alpha_{\tilde\mu\tilde\nu}(x)=\alpha_{\tilde\mu}(x)+
\alpha_{\tilde\nu}(x+\hat {\tilde\mu})-
\alpha_{\tilde\mu}(x+\hat {\tilde\nu})-\alpha_{\tilde\nu}(x)$,
$\alpha_{\tilde\mu}(x) = a A_{\tilde\mu}(x)$, 
and $a$ is the lattice spacing. For future 
reference, we also define the link matrix, 
$U_{\tilde\mu}(x) \equiv \exp[i\alpha_{\tilde\mu}(x)]$. 
The dimensionless coupling constant appearing in~\eq\nr{ab_latt_action} is 
taken to be
\be
 \beta_G(z) = \frac{\Delta(z)}{a} \;.
\ee
In the non-Abelian case, we employ the standard
Wilson action, 
\be
 S_E^{(d+1)} = \sum_z \beta_G(z) \sum_x \sum_{\tilde\mu < \tilde\nu}
 \Bigl( 1 - \frac{1}{N_c} \re \tr P_{\tilde\mu\tilde\nu} \Bigr),
\ee
where the naive continuum limit implies 
\be
 \beta_G(z) = \frac{2 N_c \Delta(z)}{a}\;.
 \la{beta}
\ee
Rather than $\beta_G(z)$, 
we will often equivalently refer to $\Delta_0/a$  
to fix the lattice spacing, where $\Delta_0 \equiv \Delta(z=0)$. 
Note that we can view $a$ as being constant throughout the lattice: in our 
case a non-constant $\beta_G(z)$ does not imply varying lattice spacing.

It should be noted that, as we have discussed in~\ref{app:cases}, 
the value of $\Delta_0$ does not affect at all the spectrum obtained
in the non-interacting limit. For a weak coupling, the criteria for
discretisation and finite volume effects to be small are simply 
\be
 a \ll \frac{1}{m_0} \ll L,T, 
\ee
where $L,T$ are the linear extents of the system
in the $r$ and $t$ directions, respectively.  
On the lattice, however,  
$\Delta_0$ determines the strength of interactions. In general, 
lattice discretisation effects are larger and the gauge theory  more
strongly coupled where $\beta_G(z)$ is smaller, if  $a m_0$ is kept
fixed. We return to this issue presently. 

It is useful to note  that if we think in terms of the mode 
decomposed action \linebreak in~\eq\nr{n_action}, then the mode $n$
can effectively 
can be assigned a 2d action at any fixed $z$, with 
\be
 \beta_G^{(\rmi{eff},n)}(z) \equiv \frac{2 N_c}{a^2 [g_n^\rmi{ext}]^2} =
 \frac{2 N_c}{a^2 \psi_n^2(z)}. 
\ee
Parameterising the dimensionless 2d link matrix $U_\mu$ as
\be
 U^{(n)}_\mu(x;z) = e^{i a \psi_n(z) T^b A^b_\mu(x)},
\ee
where $T^b$ are the Hermitean generators of SU($N_c$), 
the naive discretisation 
of the $n=0$ part of~\eq\nr{n_action} then becomes 
\be
 S_E^{(\rmi{eff})}(z) =  \beta_G^{(\rmi{eff},0)}(z)
 \sum_x \biggl[
 \sum_{\mu < \nu} \Bigl( 1 - \frac{1}{N_c} \re \tr P_{\mu\nu}^{(0)} \Bigr) +   
 (a m_0)^2 \sum_{\mu} \Bigl( 1 - \frac{1}{N_c} \re \tr U_{\mu}^{(0)} \Bigr)
 \biggr] \;. \la{2d}
\ee
An action of the form in~\eq\nr{2d}
is of course not gauge invariant, and thus in general not 
(perturbatively) renormalisable. It also does not yield
the correct naive continuum limits for the cubic and
quartic self-interactions, if $\alpha_3,\alpha_4 \neq 1$. 
Nevertheless, we might still hope~\eq\nr{2d}
to contain some qualitative features of the effective low-energy 
dynamics, to the extent that the theory is weakly coupled, 
and the results are only moderately dependent of 
the lattice spacing (or ultraviolet physics), as may indeed be 
expected to be the case in two dimensions~\cite{Bardeen:1978cz}.

The observable we measure on the lattice is 
the static force. The definitions follow 
\eqs\nr{defW}, \nr{defV}, \nr{force}, only the Wilson line is 
constructed by multiplying together the link matrices around the
rectangle, both in the non-Abelian and in the Abelian cases.  We
determine the force then as 
\be 
 F(r+\fr12 a;z) \equiv \frac{V(r+a;z) - V(r;z)}{a}. 
 \la{l_force}
\ee
Note that in the Abelian case the potential is invariant 
in $r \to L - r$ 
and the force then, for a finite $L$, takes the form 
\be
 F(r;z)=\sum_{n\geq 0} \frac{\psi_n^2(z)}{2}~\frac{\sinh{m_n(\frac{L}{2}-r)}}
 {\sinh{\frac{m_n L}{2}}}
 \;, 
 \la{pFr} \la{eq:sinh}
\ee
instead of~\eq\nr{Fr}.  
For the non-Abelian case
such a periodicity would only arise for a force
defined from the correlator of two Polyakov loops (see, e.g,~\cite{lw}). 

\section{Numerical results}
\la{se:lattice}

We now present our numerical results, obtained with standard Monte Carlo
simulation techniques. The update is a 1:4 mixture of 
heat-bath~\cite{hb,kenpen} and over-relaxation~\cite{ovr} sweeps. 
In the SU($2$) simulations, we use the following statistical noise reduction 
steps in the Wilson loop measurements:

1. First, we perform {\em link integration}~\cite{linkint} for the links 
in the $t$-direction, substituting each with the appropriate 
(and exactly calculable) local statistical average link.

2. Then, we do two {\em smearing}~\cite{smear} steps for the links along 
the $(r,z)$-plane, 
``fuzzying'' the $r$-sides of the Wilson loops by two lattice units in the 
$z$-direction, with rapidly decreasing weights.  This enhances the coupling 
to the lowest modes, which are slowly varying in $z$.

Both the link integration and the smearing must be performed taking 
into account the varying coupling $\beta_G(z)$.
We perform between $10^5$ and $4\cdot 10^5$ sweeps and collect the data 
typically in 100 bins. Errors are estimated with a standard jackknife analysis.

As our goals here are of a qualitative nature only, we should stress
that these are still very simple small scale simulations. Presumably
our numerics could have been significantly improved for instance 
by implementing the advanced methods introduced in~\cite{lw}.

\subsection{Gaussian weight function}
\la{num:gauss}

\begin{figure}[tb]

\centerline{~~\begin{minipage}[c]{7.5cm}
    \psfig{file=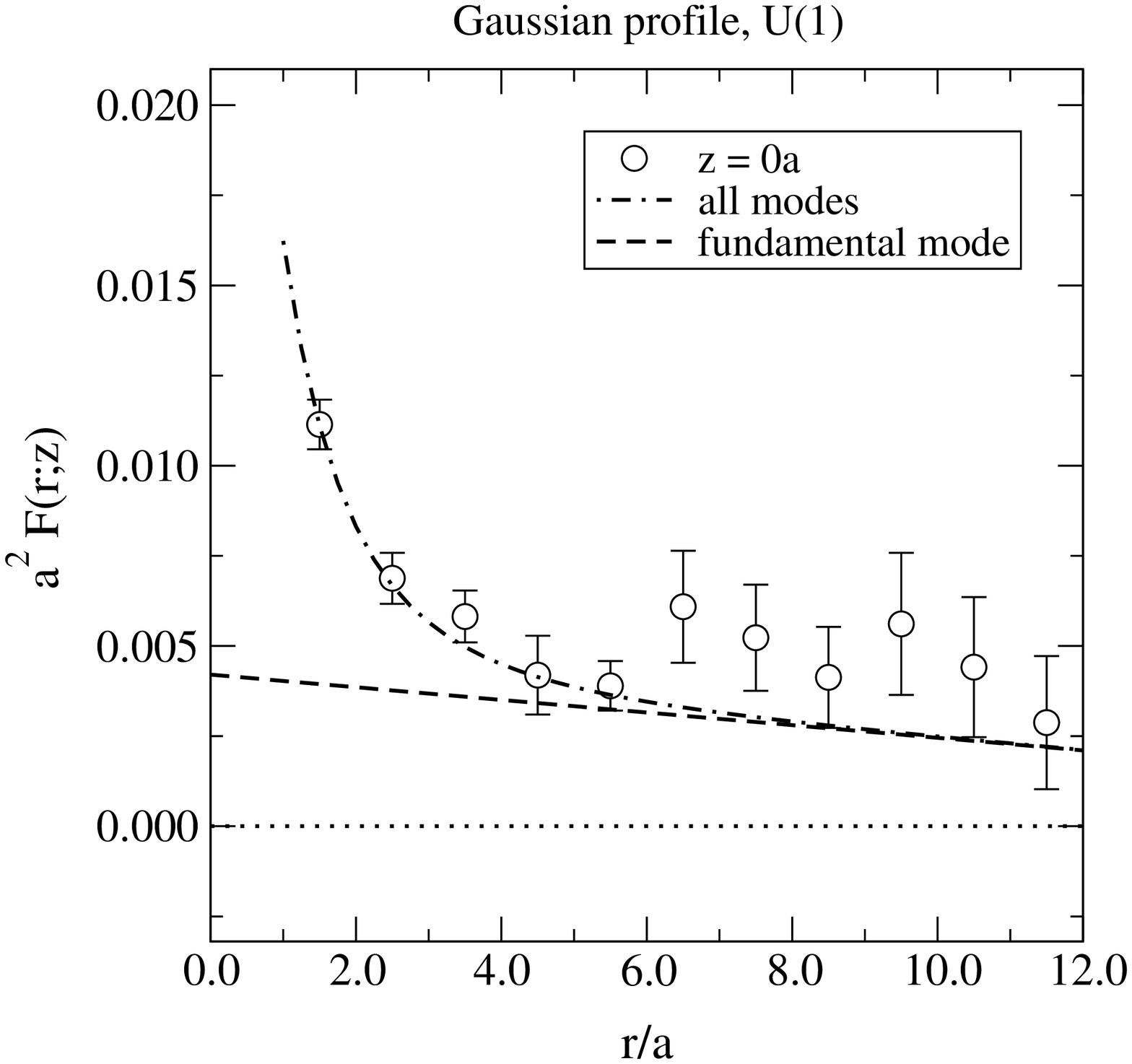,angle=0,width=7.5cm}
    \end{minipage}%
    ~~~~~\begin{minipage}[c]{7.5cm}
    \psfig{file=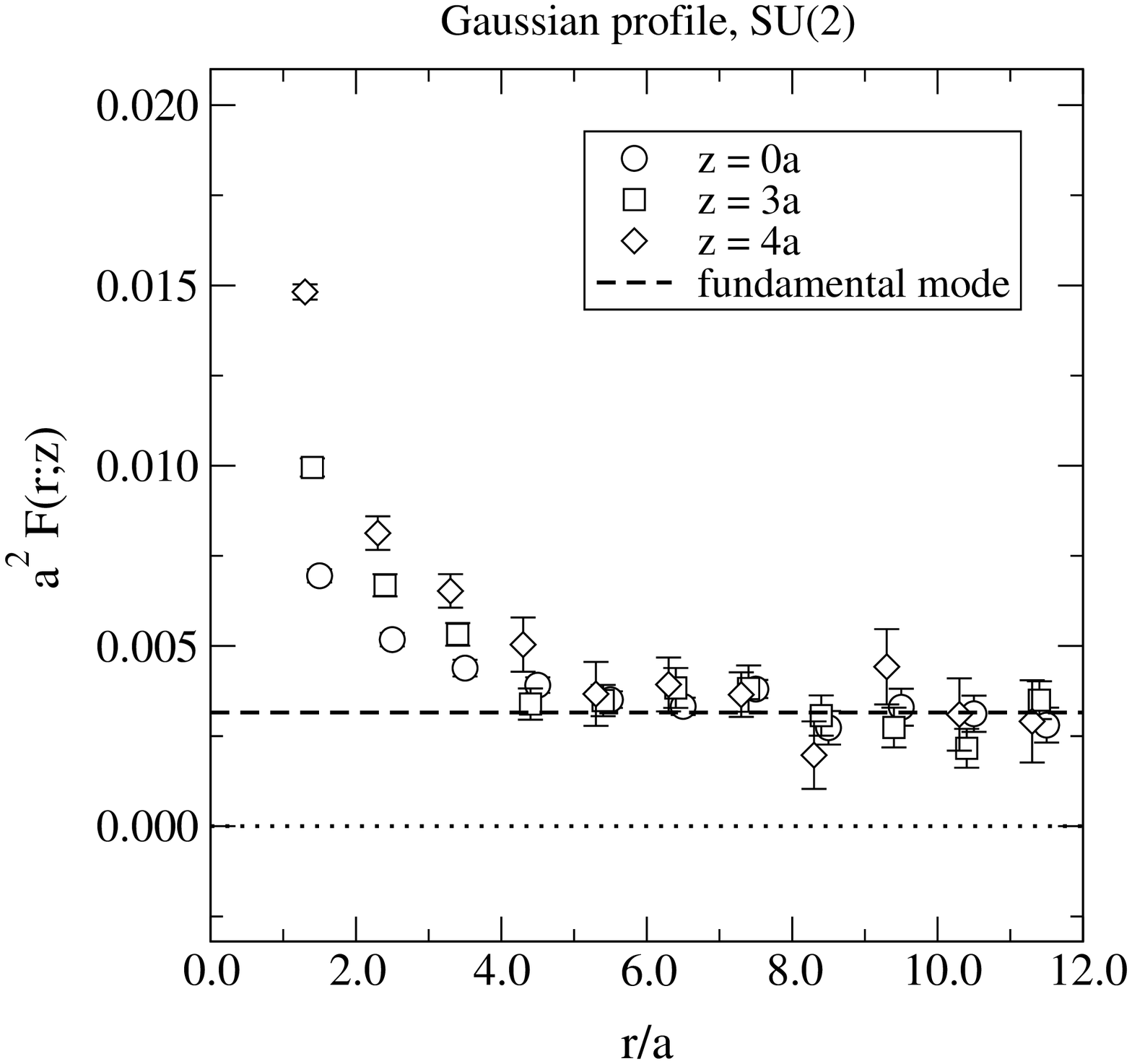,angle=0,width=7.5cm}
    \end{minipage}}

\vspace*{0.5cm}

\caption[a]{The force $F(r;z)$ for the Gaussian 
    weight function, \eq\nr{gauss}, at different fixed values of $z$, 
    in the Abelian (left; volume = $48^2\times 14$) and non-Abelian 
    cases (right; volume = $24^2\times 14$).  The perturbative values
    are also shown. The Abelian case has a finite slope because of 
    the periodicity discussed around~\eq\nr{eq:sinh}.}

\la{fig:gauss}
\end{figure}

We will start  with a study of a Gaussian weight function,
\be
 \Delta(z) \equiv \Delta_0 \exp \Bigl( -\fr12 m^2 z^2 \Bigr)\;. 
 \la{gauss}
\ee
The spectrum following from it is discussed in Appendix \ref{app:gauss},
and goes as $m_n = \sqrt{n} m$, $n=0,1,2,...\;$. The zero mode,
$\psi_c \equiv \psi_0$, does have a finite coupling
in this case, since $\int_z \Delta(z)$ is finite.  Note that 
even though the wave function $\psi_c$ 
is constant in $z$, this mode is said to have been localised, 
in the sense that $\Delta(z) \psi_c^2$ is
centered around $z=0$. 

The original theory has two parameters,  $\Delta_0,~m$.
As both of them are dimensionful ($[\Delta_0] = $GeV$^{-1}$, $[m] =
$GeV in (2+1)d),  continuum physics only depends on their product. 
Moreover, the mass spectrum, and thus the dynamics of the Abelian
theory,  are completely independent of $\Delta_0$. Correspondingly, 
it appears that the non-Abelian theory can be made weakly interacting 
by choosing a large value of $m \cdot \Delta_0$. This 
argument might fail, however, because $m_0 \cdot \Delta(z)$
is exponentially small at large~$z$.

The lattice introduces a further dimensionful parameter, the 
lattice spacing $a$. It should be chosen small enough such that
discretisation effects are harmless. Somewhat arbitrarily, we then 
fix $a, m, \Delta_0$ such that
\be
 (a m)^2 = 0.1, \quad 4 \Delta_0/a = 60.0.
\ee
Thereby the theory should be weakly coupled
($m \cdot \Delta_0 \approx 4.7$), and also close 
to continuum behaviour ($a\cdot m \approx 0.3$). 
The extent of the lattice
in the $z$-direction is chosen as $14\, a \approx 4.4 m^{-1}$
(cf.~\fig\ref{fig:gauss}).

The massless zero mode present in the system should 
dominate the physics at large distances. Its coupling is independent 
of $z$ and, according to~\eqs\nr{linear}, \nr{CA}, 
\be
 F(r;z) = \fr12 C_A \psi_c^2 = \frac{C_A}{2 \int_{z'} \Delta(z')} , 
\ee
where $C_A  = 1$ for U(1), $3/4$ for SU(2). 
The other modes give exponentially suppressed contributions,
according to~\eq\nr{Fr}. 

In fact, we can easily make an exact continuum prediction for the
full force $F(r;z)$
in the Abelian case. The analytic values of $m_n=\sqrt{n}m$
and $\psi_n(z=0)$ are given in Appendix~\ref{app:gauss},
and can be plugged into \eq(\ref{pFr}). 
The prediction is compared with numerical data 
in \fig\ref{fig:gauss} (left).
Although the data becomes noisy at large $r$, we can conclude that there is
agreement within statistical errors, confirming
that discretisation effects are under control. 

The numerical result for $F(r;z)$ in the SU($2$) case is shown
in \fig\ref{fig:gauss} (right). It indicates that the 
large distance behaviour is successfully predicted by the perturbative 
analysis. Moreover, the fact that the constant value to which 
$F(r;z)$ tends  does not depend on $z$ confirms that the constant mode 
dominates in that regime. We see also
the divergence of $F(r;z)$ at small distances,
as in the U($1$) data.
Qualitatively, the U($1$) and SU($2$)
cases yield very similar results. This is 
a peculiarity of 2d physics, however; 
had we compactified onto three dimensions, our expectation would 
be $F(r;z)\sim 1/r$ for U($1$), and yet still the confining constant 
force for SU($2$).

\subsection{Sharp weight function}
\la{num:sharp}

We next study a warped ``sharp'' weight function, 
\be
 \Delta(z) \equiv \Delta_0 
 \exp\Bigl( -M |z| + \fr12 m^2 z^2 \Bigr)\; .   
 \la{sharp}
\ee
The corresponding spectrum is discussed in Appendix~\ref{app:sharp}.

\begin{figure}[tb]

\centerline{~~\begin{minipage}[c]{7.5cm}
    \psfig{file=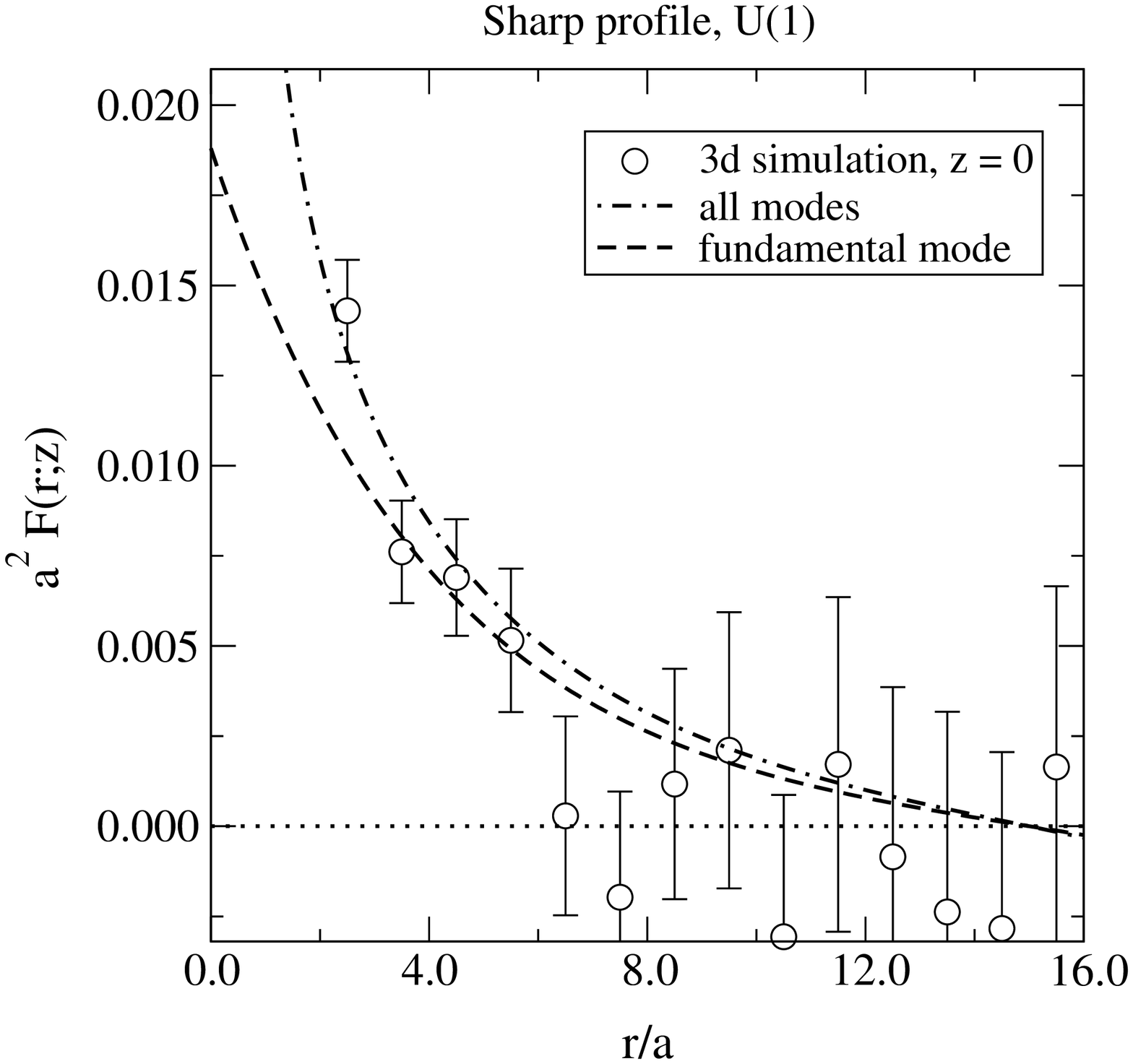,angle=0,width=7.5cm}
    \end{minipage}%
    ~~~~~\begin{minipage}[c]{7.5cm}
    \psfig{file=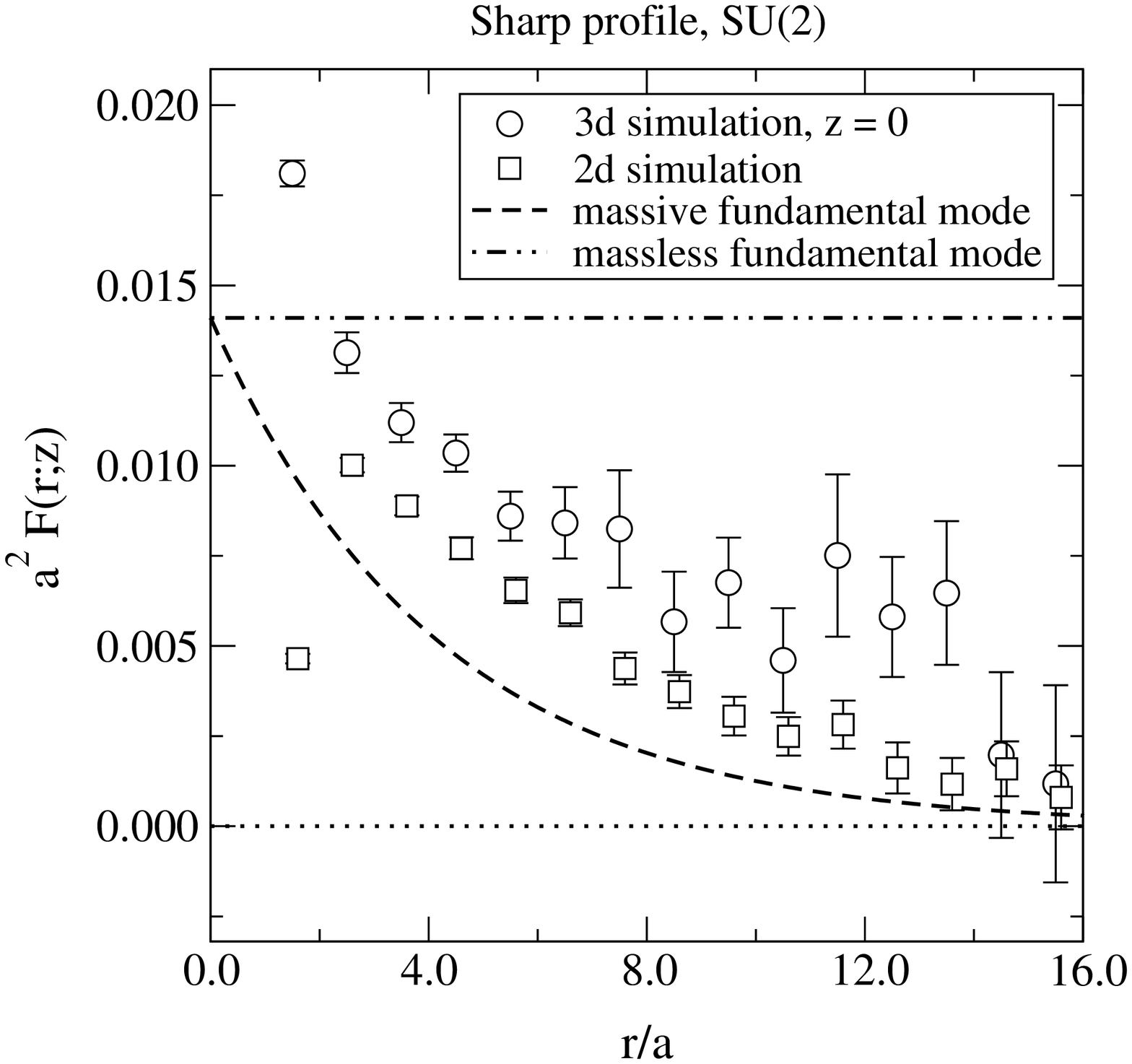,angle=0,width=7.5cm}
    \end{minipage}}

\vspace*{0.5cm}

\caption[a]{The force $F(r;z)$  
    for the sharp weight function, \eq\nr{sharp}, at $z = 0$ 
    in the Abelian (left) and non-Abelian cases (right). In the Abelian 
    case (where no noise reduction techniques were used)
    the force is anti-symmetric with respect to $r/a=15$.
    For comparison we also show the result from a 
    2d simulation based on~\eq\nr{2d}.}

\la{fig:sharp}
\end{figure}

In addition to the requirements for the Gaussian case above, 
leading to a weak effective coupling and small discretisation effects, 
we are now faced with an additional constraint, 
as well as an additional parameter allowing to satisfy it: 
we want to tune $M/m$ such that the ``fundamental mode'' $\psi_0(z)$, 
the one with the lightest non-zero mass $m_0$, is much lighter
than the next mode, with mass $m_1$. This tends to make the 
choice of parameters somewhat less transparent. 
{}From the point of view of the infinite volume setup, 
the zero mode $\psi_c$ is an artifact of the simulation, 
whose effects ought also to be kept numerically small.

In practice, we choose the parameters as
\be
 a m = 0.50, \quad
 a M = 0.75, \quad
 4 \Delta_0/a = 35.0\;.
\ee
The spectrum resulting from these parameters is discussed 
in Appendix~\ref{app:sharp}. The mass of the 
``fundamental mode'', $m_0$, is $(m_0/m)^2 = 0.235$. 
In lattice units, therefore, 
\be
 {\xi_0}/{a} \equiv ({a m_0})^{-1} \approx 4.1 \;.
\ee
The first excited state with a finite coupling $\psi_n(z)$ at $z=0$, 
on the other hand, has a correlation length $\xi_2/a \approx 1.4$.
Thus the fundamental mode should indeed dominate the infrared physics. 
For the lattice size used, $30^2\times 18$, 
the effective couplings of the zero and fundamental mode are 
\be
 a \psi_c     \approx 0.048, \quad
 a \psi_0 (0) \approx 0.194 \;. \la{psi_sharp}
\ee
We observe that because of the finite volume, 
$a \psi_c$ is not quite zero yet. 
The contributions from the different modes 
to $a^2 F(r;z)$ follow from~\eq\nr{Fr} and, fortunately, 
the effect of $a \psi_c$ turns out to be smaller than our 
error bars. Note also that a 
dimensionless 2d effective coupling can be estimated as 
\be
 \frac{\psi_0}{\pi m_0} \approx \frac{0.79}{\pi}\;,
\ee
and the parameters $\alpha_3$ and $\alpha_4$ 
defined in \eq(\ref{al}) evaluate to
\be
 \alpha_3=0.804\qquad \alpha_4=0.749,
\ee
suggesting that the perturbative picture, as well as a 2d action 
of the form in~\eq\nr{2d}, should be qualitatively applicable. 

The data is shown in~\fig\ref{fig:sharp}. 
In the Abelian case, we insert the values of $\psi_n(z=0)$ and $m_n$ 
computed in \ref{app:cases} into \eq(\ref{pFr}), to obtain the 
continuum prediction, shown with the dashed-dotted line. The dashed line 
shows the contribution of the fundamental mode alone. 
The simulation is quite consistent with the exponential decay of $F(r;z)$.
However, the periodicity in the $r$-direction makes the extraction of 
the force quite difficult and noisy when $r\approx L/2$.

In the SU($2$) case, the value of the fundamental mode prediction 
at $r=0$ sets the scale for the asymptotic confining force, were the 
fundamental mode massless (dashed-dotted line). 
Clearly, the lattice data is consistent rather
with a decaying force, or the ``breaking of the string'', on a 
distance scale given by $\xi_0 = m_0^{-1}$. As a comparison, 
we show also the result for the
2d system, \eq\nr{2d}, using the tree-level values of 
$a \psi_0(0)$, $a m_0$ as input. 
We observe the same qualitative behaviour, 
although the (2+1)-dimensional case leads to 
a stronger force at small 
distances, due to the exchange of higher modes.

Finally, we 
reiterate that because of the finite value of $a\psi_c$ 
in our finite box (cf.\ \eq\nr{psi_sharp}), the 3d force
should at very large distances still approach a finite non-vanishing 
value, $\sim 0.001$, which is however beyond our resolution.

\subsection{Smooth weight function}
\la{num:smooth}

We end by studying a ``smooth'' weight function, 
\be
 \Delta(z) = \Delta_0 
 \exp\Bigl( -\fr12 M^2 z^2 + \fr14 m^4 z^4 \Bigr)\; .   
 \la{smooth}
\ee
The corresponding spectrum is discussed in Appendix~\ref{app:smooth}.

\begin{figure}[tb]

\centerline{~~\begin{minipage}[c]{7.5cm}
    \psfig{file=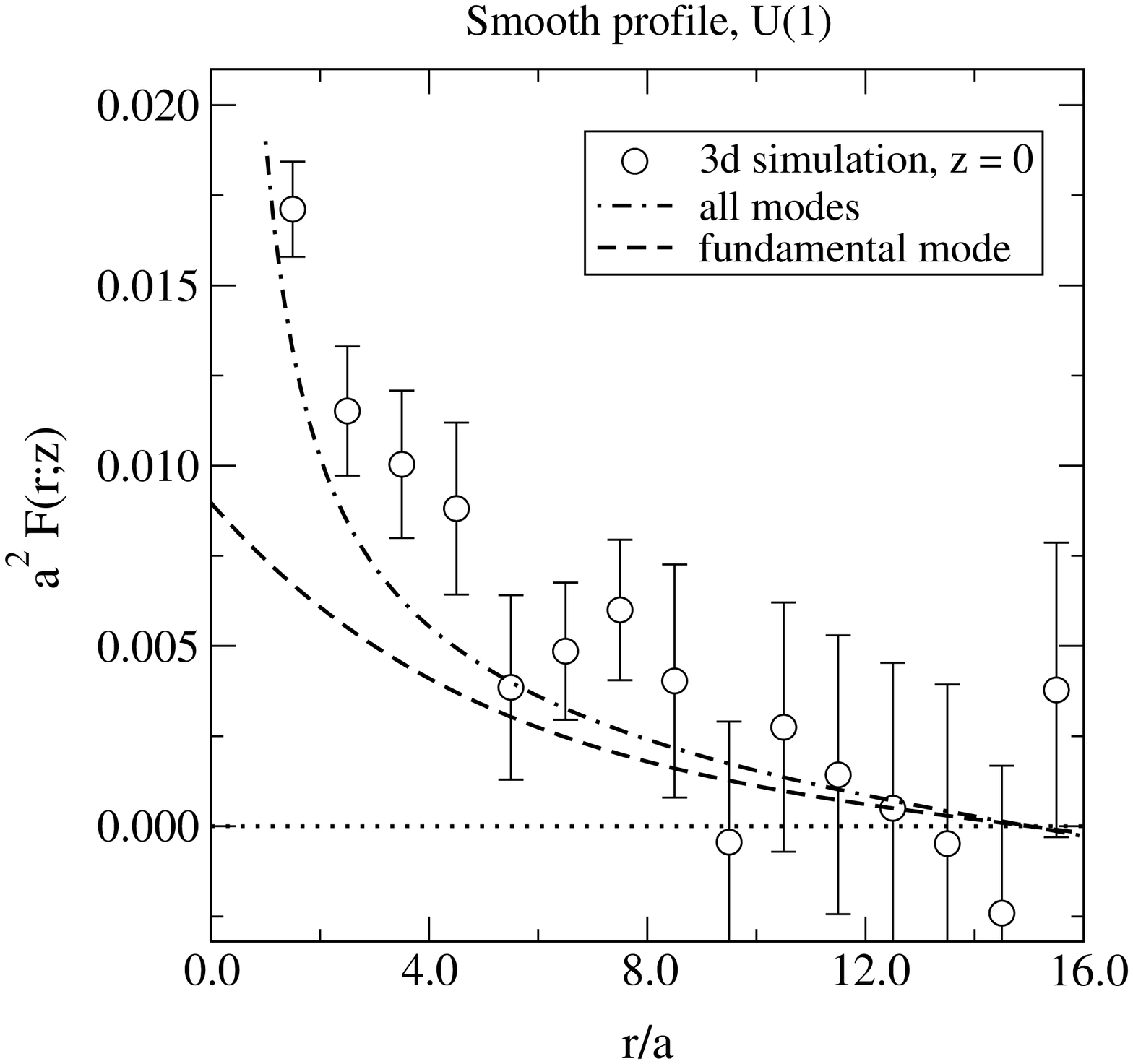,angle=0,width=7.5cm}
    \end{minipage}%
    ~~~~~\begin{minipage}[c]{7.5cm}
    \psfig{file=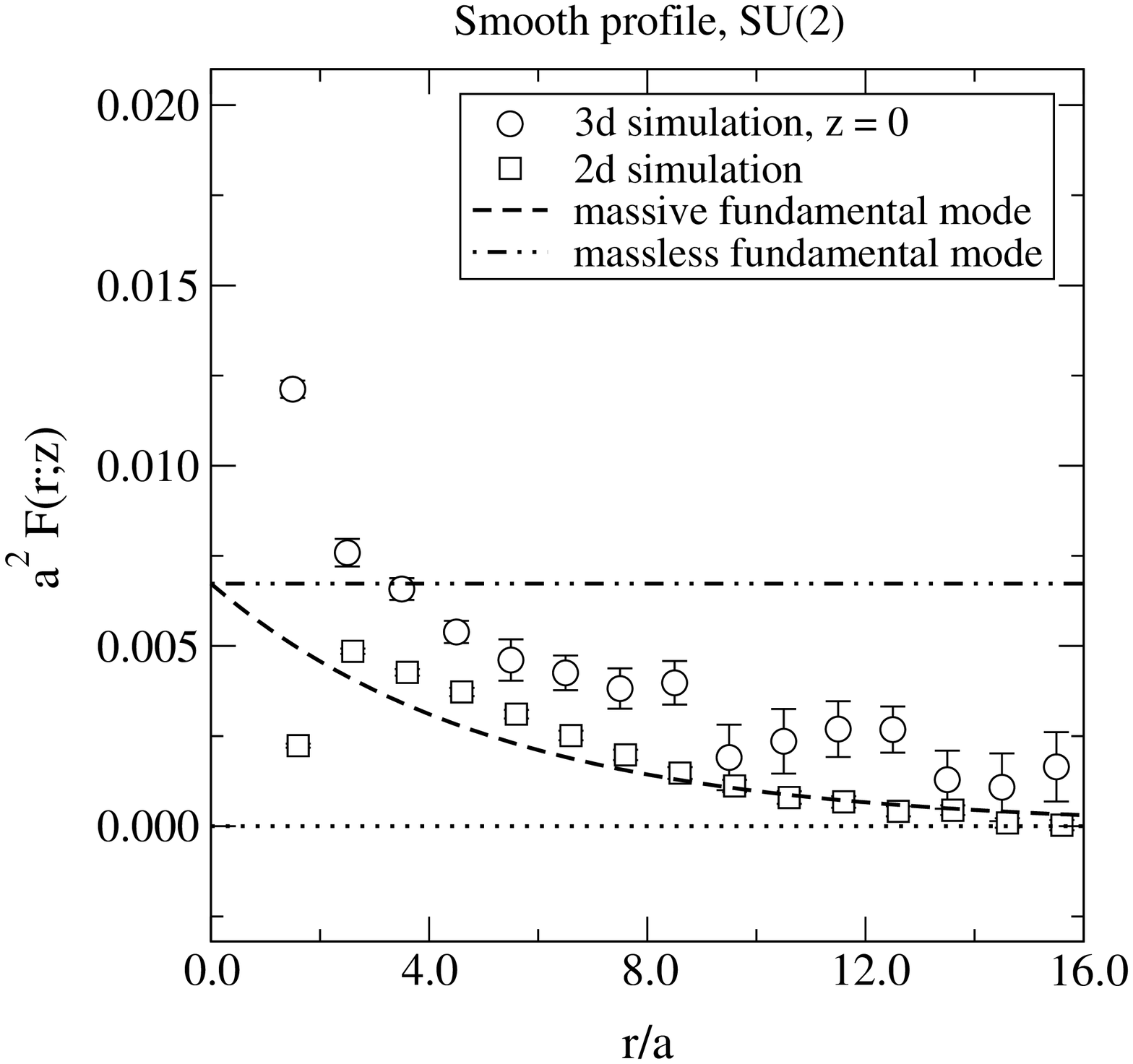,angle=0,width=7.5cm}
    \end{minipage}}

\vspace*{0.5cm}

\caption[a]{The force $F(r;z)$  
    for the smooth weight function, \eq\nr{smooth}, at $z = 0$, 
    in the Abelian (left) and non-Abelian cases (right). 
    In the Abelian case (where no noise reduction techniques were used)
    the force is anti-symmetric 
    with respect to $r/a = 15$.
    For comparison we also show the result from a 2d simulation 
    based on~\eq\nr{2d}.}

\la{fig:smooth}
\end{figure}

We choose the parameters and lattice spacing such that 
\be
 a m = 0.2778, \quad
 a M = 0.3889, \quad
 4 \Delta_0/a = 35.0\;.
\ee
According to Appendix~\ref{app:smooth}, 
the correlation length of the fundamental mode is then
\be
 \xi_0/a = (a m_0)^{-1} \approx 5.2.
\ee
The couplings of the zero and fundamental mode are, at $z=0$, 
\be
 a \psi_c    \approx  0.054, \quad
 a \psi_0(0) \approx  0.134.  \la{psi_smooth} 
\ee
These determine the string tension, according to~\eq\nr{Fr}. 
We also note that the dimensionless coupling related to the 
fundamental mode is 
\be
 \frac{\psi_0}{\pi m_0} \approx \frac{0.69}{\pi} \;, 
\ee
and the parameters $\alpha_3$ and $\alpha_4$ defined in \eq(\ref{al})
evaluate to
\be
 \alpha_3=0.705\qquad \alpha_4=0.695, 
\ee
supporting again the qualitative applicability
of the perturbative picture, as well as of the 2d action in~\eq\nr{2d}.

The lattice simulation is carried out with the volume $30^2\times 18$.
The data is shown in \fig\ref{fig:smooth}. We proceed as in the ``sharp'' 
case to compute the prediction for $F(r;z)$. 
The U($1$) data is consistent with the analytic prediction, 
although large statistical errors make this statement rather weak. 
For the SU($2$) case we again observe 
that the behaviour of the 3d simulation agrees
at large distances within statistical errors with the 
{\em massive} fundamental mode
prediction, as well as with results obtained with the 2d action 
in~\eq\nr{2d}, using the tree-level values of 
$a \psi_0(0)$, $a m_0$ as input. At smaller distances, the static force is 
stronger in the (2+1)-dimensional system, as noted previously.
Note also again that because of the finite value of $a\psi_c$ 
in our finite box (cf.~\eq\nr{psi_smooth}), the 3d force
should at very large distances still approach a finite 
non-vanishing value, $\sim 0.001$.

\section{Conclusions}
\la{se:concl}

We have studied in this paper some physical properties of a pure gauge  
field theory, living in a space where the coupling constant 
$g^2(z)\equiv \frac{1}{\Delta(z)}$ varies along one spatial 
direction. Using a mode decomposition and working in the gauge $A_z=0$,
an effective lower dimensional action was already derived in \cite{st}.
However, in the non-Abelian case, several complications arise:
all Kaluza--Klein like modes are coupled through cubic and quartic terms, 
possible non-perturbative effects in regions where the coupling is large
make the validity of the perturbative analysis unclear, and the naively 
truncated action for the low-energy sector is non-renormalisable. These
difficulties motivated a lattice simulation of the non-Abelian theory 
in (2+1) dimensions, as well as, for calibration, 
simulations of the Abelian theory in the
same background (in which case our analytic predictions 
are exact in the continuum limit).

In the case of a Gaussian profile $\Delta(z)$, our numerical data 
confirms the presence of a massless constant mode, which gives rise to 
a constant force both in the Abelian and the non-Abelian cases, in spite 
of the coupling becoming strong at large~$z$.
Note that even though the mode is constant in $z$, this mechanism
is conventionally called the localisation of massless vector bosons, 
since $\Delta(z) \times$[the mode] is sharply centered. 

For two different profiles such that $\int_z \Delta(z)=\infty$,
on the other hand, where the massless mode decouples from the theory, 
the lattice data is consistent with 
the long distance dynamics being dominated by a single 
massive localised ``fundamental'' vector mode.
Because of a large hierarchy between the mass $m_0$ of the fundamental
mode, and those of the higher modes, this 
regime can set in even at distances somewhat smaller than the correlation 
length of the fundamental mode.

Thus we confirm the qualitative picture based on perturbation theory 
in these (2+1)-dimensional systems.  It would be interesting to extend 
the study to a (3+1)-dimensional case, to check whether our conclusions 
depend on the peculiarities of the two-dimensional effective theory.

\section*{Acknowledgements}

We thank Martin L\"uscher and Peter Tinyakov for discussions.
This work was partly supported by the RTN network {\em Supersymmetry 
and the Early Universe}, EU Contract No.\ HPRN-CT-2000-00152, as well as
by the FNRS, grant No.~20-64859.01. One of us (H.M.)\
thanks the University of Lausanne for the generous
{\em Bourse de perfectionnement et de recherche}.


\appendix
\renewcommand{\thesection}{Appendix~\Alph{section}}
\renewcommand{\thesubsection}{\Alph{section}.\arabic{subsection}}
\renewcommand{\theequation}{\Alph{section}.\arabic{equation}}

\newpage

\section{Energy spectra for various weight functions}
\la{app:cases}

In this Appendix, we determine explicitly the spectra for the 
various weight functions appearing in this paper. 

As mentioned in~\cite{st}, we can write~\eq\nr{diffeq} in another
form by introducing $\chi_n(z)= \sqrt{\Delta(z)} \psi_n(z)$. Then
the eigenvalue equation takes the familiar form
\be
 - \chi_n'' + V_o(z) \chi_n = m_n^2 \chi_n, 
 \la{eigen}
\ee
where 
\be
 V_o(z) \equiv W^2(z) - W'(z), 
 \quad W(z) \equiv -\frac{\Delta'(z)}{2 \Delta(z)}.
 \la{V}
\ee
One may also introduce
\be
 V_s(z) \equiv W^2(z) + W'(z)\;;
\ee
its eigenvalues are the same as those of $V_o(z)$, except that 
one of the two has a normalisable exact zero mode, and the symmetry
properties of the two sets of wave functions with the same energy 
are the opposite~\cite{susy}. Thus, denoting $m_n^2 = E^{(n)}$ and assuming
$E_s^{(0)}=0$, we have $E_o^{(n)} = E_s^{(n+1)}, n\ge 0$.

It is useful to note that if $\Delta(z) = \Delta_0 \exp(f(z))$, then
\be
 W = - \fr12 f', \quad
 V_o = \fr14 (f')^2 + \fr12 f''\;.
\ee
Therefore, the eigenvalues are independent of $\Delta_0$.

\subsection{Gaussian weight function}
\la{app:gauss}

We start by considering 
the Gaussian weight function,~\eq\nr{gauss}, which implies that
\be
 W = \frac{1}{2} m^2 z, \quad
 V_o = \fr12 m^2 \Bigl(\fr12 m^2 z^2 -1 \Bigr) \;.
 \la{gauss2}
\ee
The eigenvalue equation, \eq\nr{eigen}, is just of the 
form of a harmonic oscillator, with shifted energy levels, 
and is immediately solved. We obtain
\be
 m_n = m \sqrt{n}, \quad n \ge 0. \la{spec_gauss}
\ee
Note 
the existence of a normalisable zero energy solution. 
It appears here for $V_o$ rather than $V_s$, 
since $\int_z \Delta(z)$ is finite. 

We also know the wave functions exactly in this case, 
\be
 \psi_n(z)=\frac{1}{2^{n/2}}~\frac{1}{\pi^{1/4}\sqrt{\Delta_0}}~
 \frac{1}{\sqrt{n!}}~H_n\Bigl(\frac{mz}{\sqrt{2}}\Bigr),
\ee
where $H_n$ are the Hermite polynomials.
Using the fact that $H_n$ have the parity of their index and that
\be 
 H_{2n}(z=0)=(-1)^n~\frac{(2n)!}{n!},
\ee
we arrive at the expression 
\be 
 \psi_{2n}(z=0)=\frac{1}{\pi^{1/4}\sqrt{\Delta_0}}\left(-\frac{1}{2}\right)^n
 \frac{\sqrt{(2n)!}}{n!}\sim 
 \frac{(-1)^n}{\sqrt{\pi \Delta_0}}~\frac{1}{n^{1/4}} \;,
\ee
where the last step is the large $n$ asymptotic behaviour. 
This means that the higher modes are not only more massive, 
but also more weakly coupled at $z=0$.

\subsection{``Sharp'' weight function}
\la{app:sharp}

\begin{figure}[tb]

\centerline{~~\begin{minipage}[c]{6.5cm}
    \psfig{file=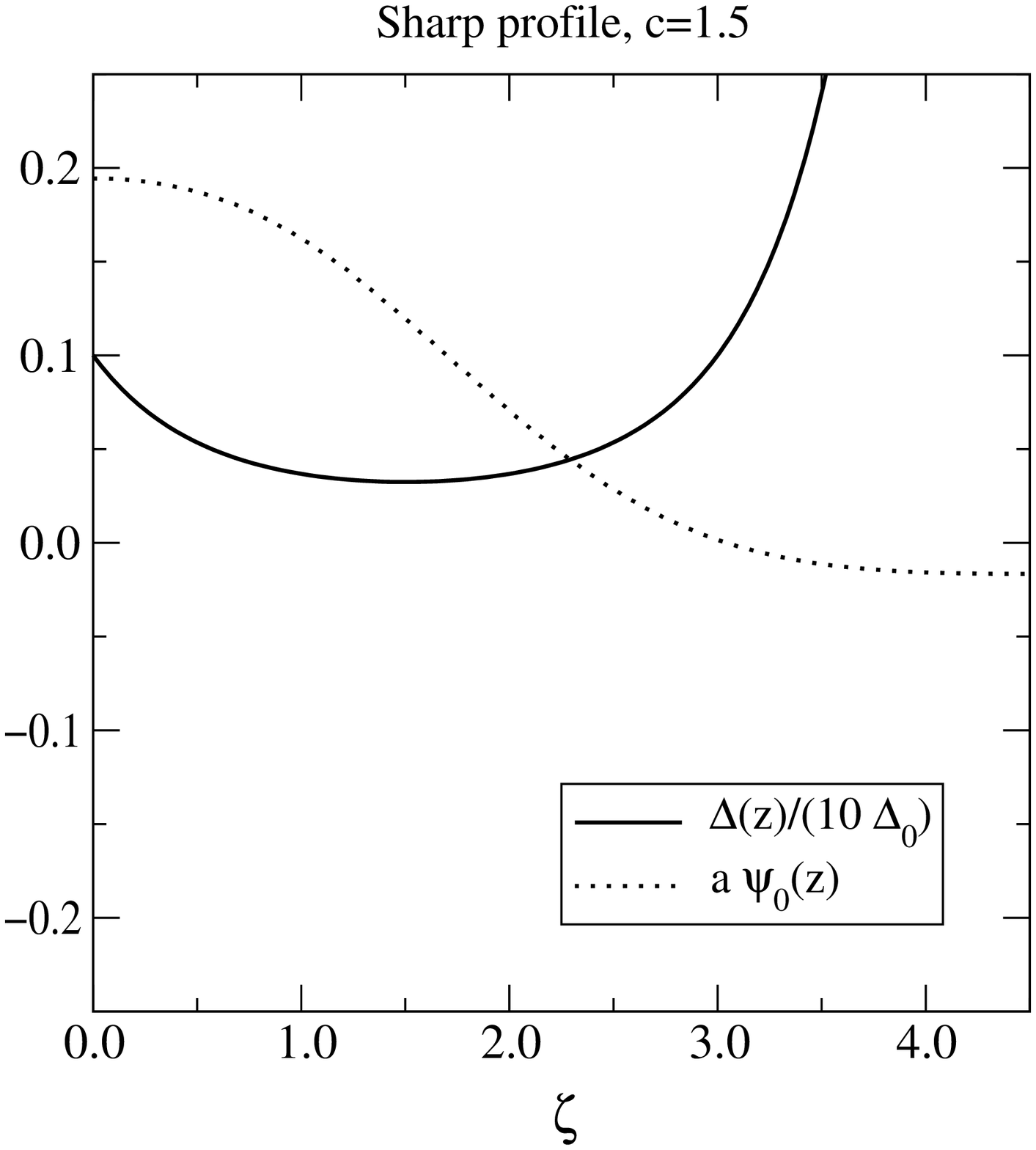,angle=0,width=6.5cm}
    \end{minipage}%
    ~~~~~\begin{minipage}[c]{6.5cm}
    \psfig{file=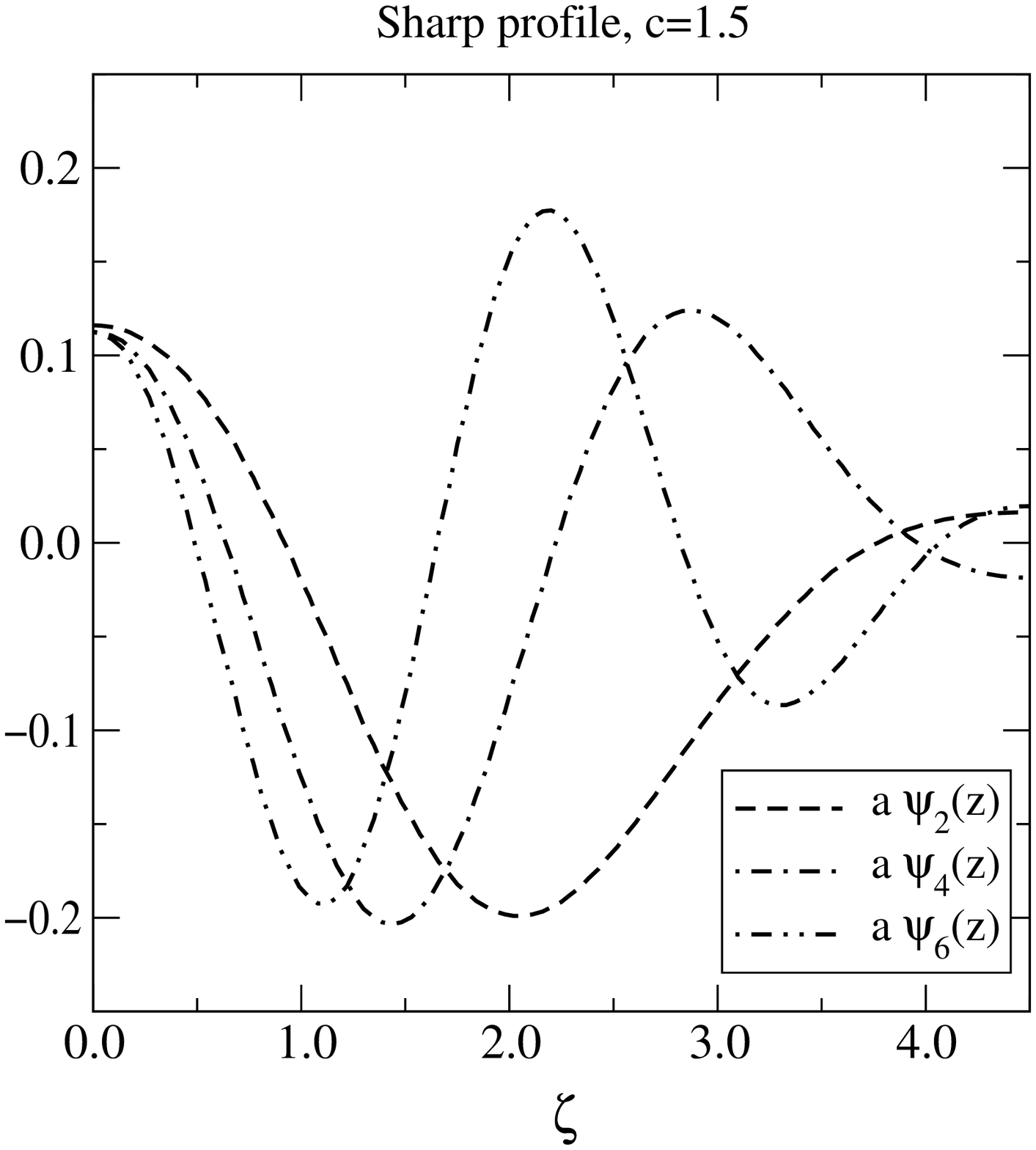,angle=0,width=6.5cm}
    \end{minipage}}

\vspace*{0.5cm}

\caption[a]{Left: The fundamental mode $\psi_0(z)$ and the profile 
     $\Delta(z)$ for the sharp wave function, 
     \eq\nr{sharp}, at $c = M/m=1.5$,  $R=4.5/m$, and 
     a specific $a$ (see Table~\ref{tab:sharp_c1.5}).
     The horizontal axis is for $\zeta = z m$. 
     Right: the first excited even modes.}

\la{fig:sharp_modes}
\end{figure}

The sharp weight function, \eq\nr{sharp}, leads to 
\be
 W = \fr12(M \mathop{\mbox{sign}}(z) - m^2 z)\; .   
 \la{sharp2}
\ee
It is convenient to 
rescale everything by $m$:
$\zeta \equiv m z$, $c \equiv M/m$, $\nu_n \equiv m_n^2/m^2$. 
\eq\nr{eigen} then becomes
\be
 -\chi_n'' + \Bigl[\fr14 (\zeta - c \mathop{\mbox{sign}}(\zeta) )^2 
 +\fr12  - c \, \delta(\zeta) \Bigr] 
 \chi_n = \nu_o^{(n)} \chi_n. 
\ee
The solutions only depend on $c$. Note that although we are using
the same notation as in~\eq\nr{eigen}, 
$\chi_n$ is here treated as a function of $\zeta$
rather than $z$.

Since the Hamiltonian is invariant under parity, one can
classify its eigenfunctions as symmetric and antisymmetric
under $\zeta \to - \zeta$. We label symmetric and antisymmetric
states with even and odd indices, respectively.

{\bf Infinite volume.}
We start by discussing the infinite volume case. 
Here we solve the problem by utilising $V_s$. Note that, 
apart from the usual relation $E_o^{(n)} = E_s^{(n+1)}$, 
$n \ge 0$, there is now the additional relation that the
eigenvalues obtained with $V_s$ and $V_o$, coming with wave
functions antisymmetric in ($z \to -z$), are trivially related 
by the addition of $m^2$, because the antisymmetric state
does not ``see'' the $\delta$ function at the origin. This 
can be expressed as $E_o^{(p)} = E_s^{(p)} + m^2$, for $p$ odd. 
Therefore, the spectrum is of the form 
\ba
 E_o^{(0)} & = &  E_s^{(1)} = \epsilon, \\
 E_o^{(1)} & = &  E_s^{(1)} + m^2 =  E_s^{(2)}, \\
 E_o^{(2)} & = &  E_s^{(3)} > E_s^{(2)}, ... \;. 
\ea
In other words, $V_s$ has a symmetric state with exactly zero energy, 
and an antisymmetric one with an exponentially small energy, $\epsilon$
(cf.~\eq\nr{expsmall} below). 
Then $V_o$ has a symmetric ground state with the energy $\epsilon$, and 
a doublet of states with a much higher energy, 
$E_s^{(3)}\approx E_s^{(2)} \gg \epsilon$. 

The explicit form of the dimensionless equation with $V_s$ becomes
\be
 -\chi_n'' + \Bigl[\fr14 (\zeta - c \mathop{\mbox{sign}}(\zeta))^2 
 -\fr12  + c \delta(\zeta) \Bigr] 
 \chi_n = \nu^{(n)}_s \chi_n. 
 \la{sharp_Vs}
\ee
Introducing the Kummer function
\be
 \phi(a;b;\zeta)=
 1+\frac{a}{b}\frac{\zeta}{1!}+\frac{a(a+1)}{b(b+1)}\frac{\zeta^2}{2!}+\dots=
 \sum_{n=0}^\infty \frac{(a)_n}{(b)_n}\frac{\zeta^n}{n!}, 
\ee
which satisfies
\be
 \zeta\frac{d^2\phi}{d\zeta^2}+(b-\zeta)\frac{d\phi}{d\zeta}-a\phi=0 \;,
\ee
the general solution of \eq\nr{sharp_Vs} reads,
for $\zeta\neq 0$ and denoting $\nu^{(n)}_s \to \nu$, 
\be
 \chi(\zeta)=
 e^{-\frac{(|\zeta|-c)^2}{4}}\left[
 A\,\phi(-\frac{\nu}{2};\frac{1}{2};\frac{(|\zeta|-c)^2}{2})+
 B\, (|\zeta|-c) \,
    \phi(\frac{1-\nu}{2};\frac{3}{2};\frac{(|\zeta|-c)^2}{2})
 \right],
\ee
where $A,B$ are constants. For $\zeta<0$, we denote $\chi$ 
by $\chi_L$, with constants $A_L,B_L$, and for $\zeta > 0$, 
$\chi_R$, with $A_R,B_R$. The symmetric wave functions 
obviously have $A_L=A_R, B_L=B_R$, the antisymmetric ones
$A_L=-A_R, B_L=-B_R$.

The boundary conditions we have to impose on the coefficients are:
\bi
\item[(a)] 
$\chi_R(0)=\chi_L(0)$,
\item[(b)] 
$\chi_R'(0)=\chi_L'(0)+c\, \chi_{L,R}(0)$,
\item[(c)] 
$\lim_{\zeta\to\infty} \chi_R(\zeta) = 
\lim_{\zeta\to-\infty} \chi_L(\zeta) = 0 \;.$
\ei
The second comes from integrating both sides of \eq\nr{sharp_Vs} 
from $-\delta$ to  $\delta$.

In both the symmetric and antisymmetric cases, the third 
condition imposes
\be
 B_R=-\sqrt{2}\frac{\Gamma(\frac{1-\nu}{2})}{\Gamma(-\frac{\nu}{2})}A_R, 
 \la{deriv}
\ee
where we used that for large $\zeta$, 
\be
 \phi(a;b;\zeta)\approx e^\zeta \zeta^{a-b}\frac{\Gamma(b)}{\Gamma(a)}.
\ee

For symmetric wave functions, the condition (a) is automatically 
satisfied. The condition (b) yields 
\be
 B_R=\frac{c\phi_1'}{\phi_3+c^2\phi_3'}A_R\;, 
\ee
where
$\phi_1'\equiv\phi'(-\frac{\nu}{2};\frac{1}{2};\frac{c^2}{2})$, 
$\phi_3\equiv\phi(\frac{1-\nu}{2};\frac{3}{2};\frac{c^2}{2})$,  
$\phi_3'\equiv\phi'(\frac{1-\nu}{2};\frac{3}{2};\frac{c^2}{2})$.
Combining this with \eq\nr{deriv}, one obtains an algebraic
equation for the energy levels: 
\be
-\nu\, c\,\phi(1-\frac{\nu}{2};\frac{3}{2};\frac{c^2}{2})+
 \sqrt{2}\frac{\Gamma(\frac{1-\nu}{2})}{\Gamma(-\frac{\nu}{2})}
 \Bigl[ \phi(\frac{1-\nu}{2};\frac{3}{2};\frac{c^2}{2})+
 c^2\frac{1-\nu}{3}
 \phi(\frac{3-\nu}{2};\frac{5}{2};\frac{c^2}{2})\Bigr] =0, 
\ee
where we made use of
\be
 \phi'(a;b;\zeta)=\frac{a}{b}\phi(a+1;b+1;\zeta) \;.
\ee
Note that since $\Gamma(-\nu/2)$ has a pole at $\nu=0$, 
$\nu=0$ is a solution for any $c$, as must be the case. 

For antisymmetric wave functions, the condition (a) yields 
\be
 B_R=\frac{\phi_1}{c\phi_3}A_R \;, 
\ee
with the same notation as above. 
The condition (b) is then automatically satisfied. 
Together with \eq\nr{deriv}, 
one again obtains an algebraic equation for the energy levels: 
\be 
\Gamma(-\frac{\nu}{2})\phi(-\frac{\nu}{2};\frac{1}{2};\frac{c^2}{2})
 +\sqrt{2}\,c\,
 \Gamma(\frac{1-\nu}{2})\phi(\frac{1-\nu}{2};\frac{3}{2};\frac{c^2}{2})=0\;.
\ee
In the limit $c \gg 1$, the approximate solution for the lowest energy
level is obtained by setting $\nu \to 0$ in the argument of the 
first $\phi$ appearing, whereby $\phi(0;\fr12;\fr{c^2}2)=1$; 
we then get
\be
 \nu_o^{(0)} = \nu_s^{(1)} 
 \approx \sqrt{{2}/{\pi}}\,c \,e^{-\frac{c^2}{2}}. \la{expsmall}
\ee
Some numerical values for $\nu$
(expressed as $\nu^{(n)}_o = \nu^{(n+1)}_s$) 
are included in~Table~\ref{tab:sharp_c1.5}.

{\bf Finite volume.}
Let us now consider the same system, but in a box with periodic
boundary conditions at $\zeta = \pm R/m \equiv \pm \hat R$, 
rather than in infinite volume:
\be
 \chi_L(-\hat R) = \chi_R(\hat R). 
\ee
Then the spectrum changes. We will study this system directly 
in terms of $V_o$, rather than $V_s$. The general form of the 
solution in terms of the Kummer functions remains the same, 
except that the Hamiltonian at $\zeta\neq 0$ has changed by 
a constant. 

We will now need to impose a boundary 
condition at $\zeta = \hat R$. Integrating the 
equation of motion for $\psi$  from $\hat R-\delta$ to $\hat R+\delta$, 
it is easily seen that $\psi'$ must be continuous 
at $\zeta=\hat R$ (because $\Delta$ is). 
This translates into 
\be
 \chi_R'(\hat R)+\hat W(\hat R)\chi_R(\hat R)=0 \;, \label{pbc}
\ee
where $\hat W \equiv W/m$.
This means that $\chi$'s derivative is not continuous at the boundary, 
which is due to the discontinuity of $\Delta'$ at $\zeta=\hat R$. 
The other boundary condition for the symmetric states 
is~\eq\nr{pbc} applied at the origin or, 
equivalently, condition (b) above, 
\be
 \chi_R'(0) + \hat W(0) \chi_R(0)=0 \;. \la{obc}
\ee
Note that \eqs\nr{pbc}, \nr{obc} imply that $\psi_R'(0) = \psi_R'(R)=0$.
In the antisymmetric case, the complete boundary conditions are
\be
 \chi_R(0) = 0, \qquad
 \chi_R(\hat R) = 0, \la{asbc}
\ee
and~\eq\nr{pbc} then imposes that $\chi'_R(\hat R)$ vanish as well. 

For symmetric wave functions, \eq\nr{obc} implies 
\be
 A_R=D B_R, \qquad
 D \equiv \frac{(1-c^2)\phi_3+c^2 \phi_3'}{c(\phi_1'-\phi_1)},
\ee
where the notation is as above. Inserting into \eq\nr{pbc},
\be
 \Bigl[1 - (\hat R - c)^2\Bigr]\phi_3^{\hat R} + 
 (\hat R - c)^2 \phi_3'^{\hat R}+ 
 (\hat R - c) (\phi_1'^{\hat R} - \phi_1^{\hat R}) D = 0, 
\ee
where 
$\phi_1^{\hat R}\equiv\phi(-\frac{\nu}{2};\frac{1}{2};
\frac{(\hat R - c)^2}{2})$, 
$\phi_1'^{\hat R}\equiv\phi'(-\frac{\nu}{2};\frac{1}{2};
\frac{(\hat R - c)^2}{2})$, 
$\phi_3^{\hat R} \equiv\phi(\frac{1-\nu}{2};\frac{3}{2};
\frac{(\hat R - c)^2}{2})$,  
$\phi_3'^{\hat R}\equiv\phi'(\frac{1-\nu}{2};\frac{3}{2};
\frac{(\hat R - c)^2}{2})$.
This equation determines $\nu$, after which 
the actual energy level is found by adding 1.

For antisymmetric wave functions, \eqs\nr{asbc} 
imply that 
\ba
 B_R=\frac{\phi_1}{c\phi_3}A_R, \qquad
 B_R=\frac{\phi_1^{\hat R}}{(c - \hat R)\phi_3^{\hat R}}A_R\;.
\ea
These are compatible only if
\be
 (c - \hat R)\phi_1 \phi_3^{\hat R}=c\phi_3\phi_1^{\hat R} \;,
\ee
which determines the eigenvalues. 

\begin{table}[t]
\begin{center}
\begin{tabular}{ccccccc}
\hline
 & \multicolumn{3}{c}{sharp profile, $c=1.5$}
 & \multicolumn{3}{c}{sharp profile, $c=2.5$} \\
 & \multicolumn{2}{c}{$\hat R = 4.5$} 
 & \multicolumn{1}{c}{$\hat R = \infty$} 
 & \multicolumn{2}{c}{$\hat R = 4.5$} 
 & \multicolumn{1}{c}{$\hat R = \infty$} \\
 $n$ & $\nu_o^{(n)}$ & $a \psi_n(0)$ & $\nu_o^{(n)}$
 & $\nu_o^{(n)}$ & $a \psi_n(0)$ & $\nu_o^{(n)}$ \\
\hline
 $c$  & 0.000 & 0.048 & ---   & 0.000 & 0.201 & ---   \\  
 $0$  & 0.235 & 0.194 & 0.209 & 0.154 & 0.148 & 0.038 \\  
 $1$  & 1.235 & 0.000 & 1.209 & 1.154 & 0.000 & 1.038 \\  
 $2$  & 1.945 & 0.116 & 1.643 & 1.815 & 0.101 & 1.191 \\  
 $3$  & 2.945 & 0.000 & 2.643 & 2.815 & 0.000 & 2.191 \\  
 $4$  & 4.432 & 0.113 & 3.201 & 4.298 & 0.106 & 2.478 \\  
 $5$  & 5.432 & 0.000 & 4.201 & 5.298 & 0.000 & 3.478 \\  
 $6$  & 7.854 & 0.112 & 4.829 & 7.723 & 0.109 & 3.867 \\  
 $7$  & 8.854 & 0.000 & 5.829 & 8.723 & 0.000 & 4.867 \\  
 $8$  & 12.24 & 0.112 & 6.504 & 12.11 & 0.110 & 5.327 \\  
 $9$  & 13.24 & 0.000 & 7.504 & 13.11 & 0.000 & 6.327 \\  
\hline
\end{tabular}
\end{center}

\caption[a]{Eigenvalues for the sharp profile. 
For the wave functions at origin one needs also
the values of $am, \Delta_0/a$, entering as 
$a \psi_n(0) \propto (a m)^{1/2} (\Delta_0/a)^{-1/2}$, 
if $c,\hat R$ are kept fixed; 
we have assumed $a m =0.5, \Delta_0/a = 8.75$,
for both values of $c$.}
\la{tab:sharp_c1.5}
\end{table}

As an example, some numerical values are given 
in Table~\ref{tab:sharp_c1.5}.
The doublet structure as discussed above is visible
in the infinite volume results (levels 1 \& 2; 3 \& 4; ...).
It can also be observed that for small $n$, the spectrum is roughly
linear in $n$ (like in \eq\nr{spec_gauss} for $m_n^2$), 
while for large values within a finite box, it starts to resemble 
more the corresponding spectrum 
$\sim n^2$. 

\subsection{``Smooth'' weight function}
\la{app:smooth}

\begin{figure}[tb]

\centerline{~~\begin{minipage}[c]{6.5cm}
    \psfig{file=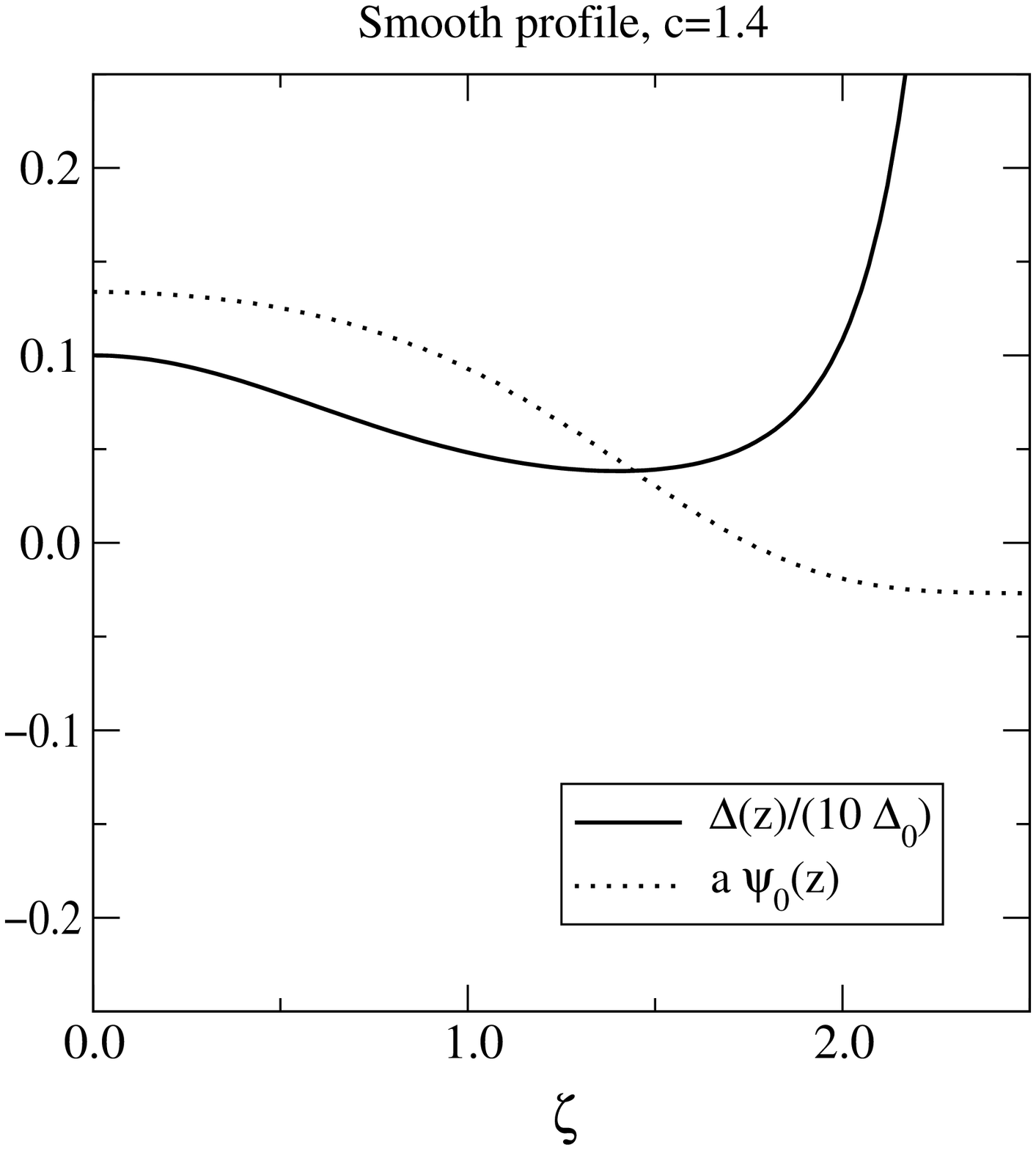,angle=0,width=6.5cm}
    \end{minipage}%
    ~~~~~\begin{minipage}[c]{6.5cm}
    \psfig{file=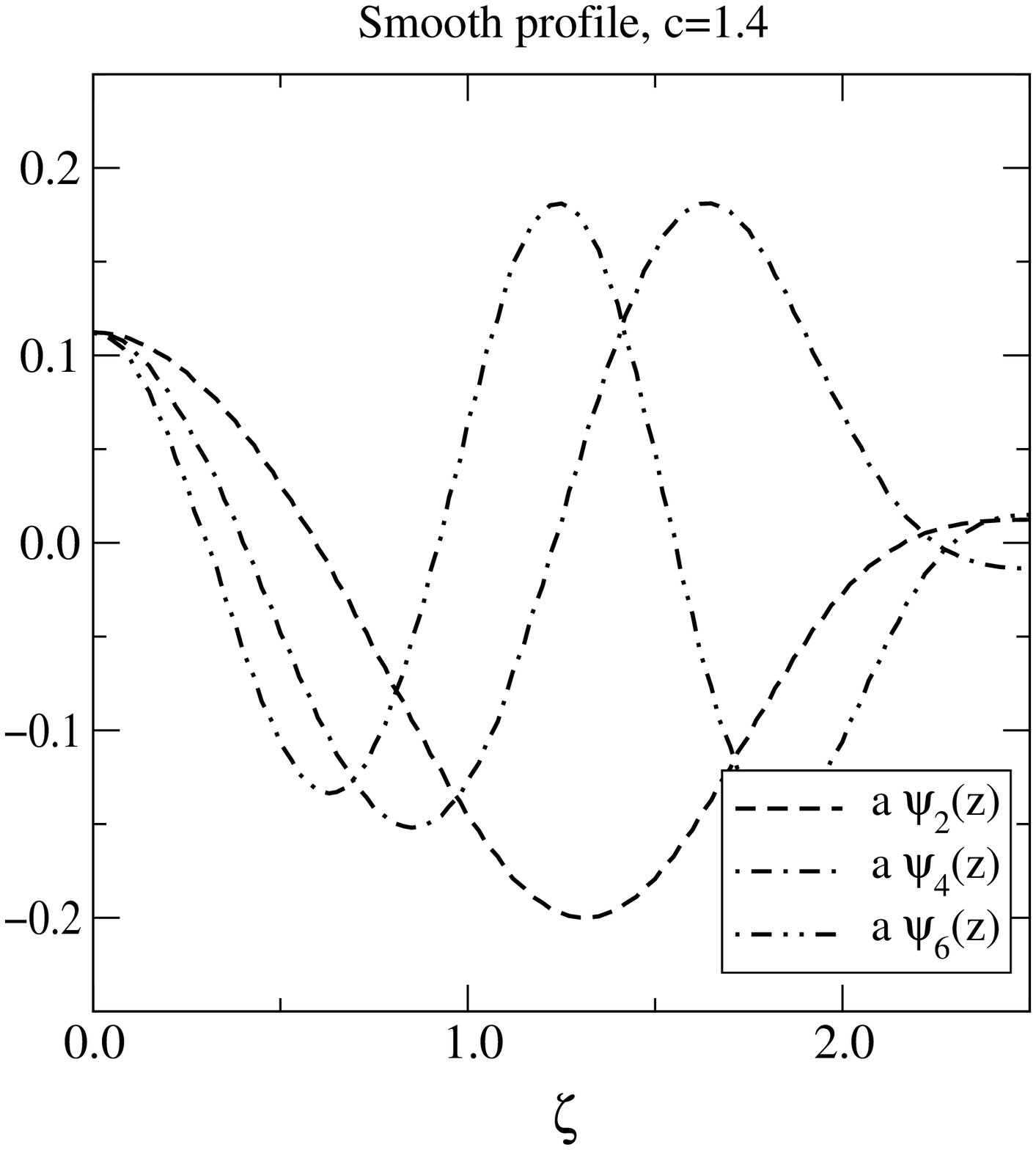,angle=0,width=6.5cm}
    \end{minipage}}

\vspace*{0.5cm}

\caption[a]{Left: The fundamental mode $\psi_0(z)$ and the profile
     $\Delta(z)$ for the smooth wave
     function, \eq\nr{smooth}, at $c = M/m = 1.4$, $R = 2.5/m$, 
     and a specific $a$ (see Table~\ref{tab:smooth_c1.4}).
     The horizontal axis is for $\zeta = z m$. 
     Right: the first excited even modes.}

\la{fig:smooth_modes}
\end{figure}

The smooth weight function,~\eq\nr{smooth}, leads to 
\be
 W = \fr12(M^2 z - m^4 z^3)\; .   
 \la{smooth2}
\ee
We rescale again 
everything by $m$: $\zeta= m z$, $c = M/m$, $\nu_n = m_n^2/m^2$. 
\eq\nr{eigen} then becomes
\be
 -\chi_n'' + \Bigl[\fr14 \zeta^2 (c^2 - \zeta^2)^2 - \fr12 c^2 + 
  \fr32 \zeta^2 \Bigr] 
  \chi_n = \nu_o^{(n)} \chi_n. 
\ee
The solutions thus only depend on $c$. 
The boundary conditions are as in~\eqs\nr{pbc}, \nr{obc}, \nr{asbc}, 
for the symmetric and antisymmetric cases, respectively.

\begin{table}[t]
\begin{center}
\begin{tabular}{ccccccc}
\hline
 & \multicolumn{3}{c}{smooth profile, $c=1.4$}
 & \multicolumn{3}{c}{smooth profile, $c=2.0$} \\
 & \multicolumn{2}{c}{$\hat R = 2.5$} 
 & \multicolumn{1}{c}{$\hat R = \infty$} 
 & \multicolumn{2}{c}{$\hat R = 3.15$} 
 & \multicolumn{1}{c}{$\hat R = \infty$} \\
 $n$ & $\nu_o^{(n)}$ & $a \psi_n(0)$ & $\nu_o^{(n)}$ 
 & $\nu_o^{(n)}$ & $a \psi_n(0)$ & $\nu_o^{(n)}$ \\
\hline
 $c$  & 0.000 & 0.054 & ---   & 0.000 & 0.034 & ---   \\  
 $0$  & 0.484 & 0.134 & 0.388 & 0.033 & 0.112 & 0.030 \\  
 $1$  & 3.020 & 0.000 & 3.016 & 3.310 & 0.000 & 3.310 \\  
 $2$  & 6.154 & 0.112 & 5.862 & 5.517 & 0.065 & 5.513 \\  
 $3$  & 9.430 & 0.000 & 9.373 & 7.912 & 0.000 & 7.912 \\  
 $4$  & 14.48 & 0.112 & 13.50 & 11.16 & 0.075 & 11.15 \\  
 $5$  & 18.51 & 0.000 & 18.17 & 14.95 & 0.000 & 14.95 \\  
 $6$  & 25.83 & 0.112 & 23.32 & 19.27 & 0.075 & 19.21 \\  
 $7$  & 30.17 & 0.000 & 28.92 & 23.92 & 0.000 & 23.92 \\  
 $8$  & 40.20 & 0.112 & 34.95 & 29.20 & 0.074 & 29.03 \\  
 $9$  & 44.63 & 0.000 & 41.37 & 34.57 & 0.000 & 34.54 \\  
\hline
\end{tabular}
\end{center}

\caption[a]{Eigenvalues for the smooth profile.
For the wave functions at origin one needs also
the values of $am, \Delta_0/a$, entering as 
$a \psi_n(0) \propto (a m)^{1/2} (\Delta_0/a)^{-1/2}$, 
if $c,\hat R$ are kept fixed; 
we have assumed $a m =5.0/18.0, \Delta_0/a = 8.75$, for $c=1.4$;
and $a m =0.45, \Delta_0/a = 25.0$, for $c=2.0$.}
\la{tab:smooth_c1.4}
\end{table}

This time we have only solved the 
eigenvalue problem numerically. The profile
$\Delta(z)$, the fundamental wave function $\psi_0(z)$, as well as the first 
excited symmetric wave functions, are shown in~\fig\ref{fig:smooth_modes}, 
for $c = M/m= 1.4$. Examples of eigenvalues are given in
Table~\ref{tab:smooth_c1.4}.

\section{Gauge invariant correlators in the Abelian case}
\la{app:correlations}

In the Abelian theory, 
\eq\nr{action},
the field strength tensor $F_{\mu\nu}$ is 
gauge invariant. This allows one to measure directly various
gauge invariant correlation functions displaying the 
essential features of the spectrum $\{ m_n \}$, as we will show
with a specific example. These correlators are however 
not available in the non-Abelian case. Other
possibilities exist, but they contain either composite operators, making
a qualitative distinction between genuinely confining and Higgs-like phases
difficult, or non-local operators, making the analysis of 
ultraviolet divergences as well as practical measurements hard.

Let us define
\ba
 {\cal O}_p(x;z) & = &  \frac{1}{T} \int\! {\rm d}t\, 
 F_{xt} (x,t,z) e^{i p t}, \\
 {\cal O}^{(n)}_p(x) & = & 
 \int_z  \Delta(z)  \psi_n(z)  {\cal O}_p(x;z) \;.
\ea
We might then consider, e.g., a weighted average over the $z$-direction, 
\be
 G^{(\rmi{all})}_p(r) \equiv \int_z \Delta(z) 
 \int\! {\rm d}x\, {\cal O}_p(x;z) [{\cal O}_p (x+r;z)]^*,
\ee
or, alternatively, a correlator only getting a contribution from 
some specified mode, 
\ba
 G^{(n)}_p(r) & \equiv & 
 \int\! {\rm d}x\, {\cal O}^{(n)}_p(x) [{\cal O}^{(n)}_p(x+r)]^*.
\ea
Employing the unitary gauge propagator
($\vec{p}, \vec{x}, \vec{x}'$ are 2d vectors)
\be
 \langle F^m_{xt}(\vec{x})F^n_{xt}(\vec{x}')\rangle=   
 \int \frac{{\rm d}^2p}{(2\pi)^2}
 \frac{p^2}{p^2+m_n^2} \delta^{mn} e^{i\vec{p}\cdot (\vec{x}-\vec{x}')},
\ee
one easily finds
\ba
 G^{(\rmi{all})}_p(r) & = &  \sum_n  G^{(n)}_p(r), \\ 
 G^{(n)}_p(r) & = & 
 \delta(r) - \frac{m_n^2}{2\sqrt{p^2 + m_n^2}} e^{-|r| \sqrt{p^2 + m_n^2}} \;. 
\ea
Therefore, the spectrum $\{m_n\}$ again manifests itself in 
the form of the exponential decay. 
Note that in contrast to the force $F(r;z)$, 
however, the wave functions $\psi_n(z)$ do not appear in these predictions.

\section{The SU($N_c$) string tension in 2d}
\la{app:tension}

In this section we recall briefly the results for the static 
potential of pure SU($N_c$) gauge theory in two dimensions. 
For a more detailed discussion see, e.g., ref.~\cite{b}.

The result for the Abelian case is shown in~\eq\nr{lo_pot}, 
with $\sum_n \psi_n^2(z)\to 1$. 
In the non-Abelian case, that result is at leading order
simply multiplied by an additional factor 
$C_A = (N_c^2 - 1)/(2 N_c)$, coming from the sum over the Hermitian
generators of the fundamental representation, $\sum_a T^a T^a$.
If we define $C_A = 1$ for U(1), the leading order
result is then as written in~\eqs\nr{Fr}, \nr{CA}.   

It remains to show that there are no higher order corrections
in the non-Abelian case. The naive argument goes as follows. 
As the potential 
is gauge fixing independent by construction,
we can choose the gauge $A_t = 0$. But then the cubic
and quartic interactions vanish, so that indeed no further
corrections should arise. 


\end{document}